\documentclass[conference]{IEEEtran}
\usepackage{nicefrac}
\usepackage{siunitx}
\usepackage{array,framed}
\usepackage{flushend}

\usepackage{microtype}
\renewcommand{\baselinestretch}{0.97}
\renewcommand{\baselinestretch}{0.982}
\usepackage{setspace}
\usepackage{booktabs}

\usepackage{tikz}

\usepackage{amsmath}
\usepackage{listings}
\usepackage{xspace}

\usepackage{hyperref}
\usepackage{bigstrut}

\usepackage{filecontents}
\usepackage{booktabs}
\usepackage{savesym}

\usepackage{changes}

\usepackage{subcaption}
\usepackage{url}
\usepackage{amsfonts}

\usepackage{amsthm}

\usepackage{todonotes}
\setuptodonotes{fancyline, inline, color=blue!30}
\definechangesauthor[name=arthur, color=red]{arthur}

\usepackage{multirow}
\usepackage[ruled,vlined]{algorithm2e}
\usepackage{algpseudocode}
\usepackage{graphicx}
\usepackage{enumitem}
\usepackage{numprint}
\usepackage[normalem]{ulem}
\usepackage{tcolorbox}
\usepackage{xurl}

\usepackage{caption}
\usepackage{subcaption}
\usepackage{diagbox}
\usepackage{xstring}
\usepackage[makeroom]{cancel}
\usepackage{calc}
\newsavebox\CBox
\newcommand\hcancel[2][0.5pt]{%
  \ifmmode\sbox\CBox{$#2$}\else\sbox\CBox{#2}\fi%
  \makebox[0pt][l]{\usebox\CBox}%
  \rule[0.5\ht\CBox-#1/2]{\wd\CBox}{#1}}

\definecolor{lightgray}{rgb}{.9,.9,.9}
\definecolor{darkgray}{rgb}{.4,.4,.4}
\definecolor{purple}{rgb}{0.65, 0.12, 0.82}

\lstdefinelanguage{JavaScript}{
  keywords={typeof, new, true, false, catch, function, return, null, catch, switch, var, if, in, while, do, else, case, break},
  keywordstyle=\color{blue}\bfseries,
  ndkeywords={class, export, boolean, throw, implements, import, this},
  ndkeywordstyle=\color{darkgray}\bfseries,
  identifierstyle=\color{black},
  sensitive=false,
  comment=[l]{//},
  morecomment=[s]{/*}{*/},
  commentstyle=\color{purple}\ttfamily,
  stringstyle=\color{red}\ttfamily,
  morestring=[b]',
  morestring=[b]"
}

\definecolor{verylightgray}{rgb}{.97,.97,.97}

\lstdefinelanguage{Solidity}{
	keywords=[1]{anonymous, assembly, assert, balance, break, call, callcode, case, catch, class, constant, continue, constructor, contract, debugger, default, delegatecall, delete, do, else, emit, event, experimental, export, external, false, finally, for, function, gas, if, implements, import, in, indexed, instanceof, interface, internal, is, length, library, log0, log1, log2, log3, log4, memory, modifier, new, payable, pragma, private, protected, public, pure, push, require, return, returns, revert, selfdestruct, send, solidity, storage, struct, suicide, super, switch, then, this, throw, transfer, true, try, typeof, using, value, view, while, with, addmod, ecrecover, keccak256, mulmod, ripemd160, sha256, sha3}, 
	keywordstyle=[1]\color{blue}\bfseries,
	keywords=[2]{address, bool, byte, bytes, bytes1, bytes2, bytes3, bytes4, bytes5, bytes6, bytes7, bytes8, bytes9, bytes10, bytes11, bytes12, bytes13, bytes14, bytes15, bytes16, bytes17, bytes18, bytes19, bytes20, bytes21, bytes22, bytes23, bytes24, bytes25, bytes26, bytes27, bytes28, bytes29, bytes30, bytes31, bytes32, enum, int, int8, int16, int24, int32, int40, int48, int56, int64, int72, int80, int88, int96, int104, int112, int120, int128, int136, int144, int152, int160, int168, int176, int184, int192, int200, int208, int216, int224, int232, int240, int248, int256, mapping, string, uint, uint8, uint16, uint24, uint32, uint40, uint48, uint56, uint64, uint72, uint80, uint88, uint96, uint104, uint112, uint120, uint128, uint136, uint144, uint152, uint160, uint168, uint176, uint184, uint192, uint200, uint208, uint216, uint224, uint232, uint240, uint248, uint256, var, void, ether, finney, szabo, wei, days, hours, minutes, seconds, weeks, years},	
	keywordstyle=[2]\color{teal}\bfseries,
	keywords=[3]{block, blockhash, coinbase, difficulty, gaslimit, number, timestamp, msg, data, gas, sender, sig, value, now, tx, gasprice, origin},	
	keywordstyle=[3]\color{violet}\bfseries,
	identifierstyle=\color{black},
	sensitive=false,
	comment=[l]{//},
	morecomment=[s]{/*}{*/},
	commentstyle=\color{gray}\ttfamily,
	stringstyle=\color{red}\ttfamily,
	morestring=[b]',
	morestring=[b]"
}

\DeclareMathDelimiter{(}{\mathopen} {operators}{"28}{largesymbols}{"00}
\DeclareMathDelimiter{)}{\mathclose}{operators}{"29}{largesymbols}{"01}

\setitemize{noitemsep,topsep=0pt,parsep=0pt,partopsep=0pt}
\setdescription{noitemsep,topsep=0pt,parsep=0pt,partopsep=0pt}

\makeatletter
\g@addto@macro{\normalsize}{%
    \setlength{\abovedisplayskip}{5pt}
    \setlength{\abovedisplayshortskip}{5pt}
    \setlength{\belowdisplayskip}{5pt}
    \setlength{\belowdisplayshortskip}{5pt}}
\makeatother

\newcommand{\etal}{\textit{et al.\ }}
\npthousandsep{,}

\usepackage{tikz}

\newcommand{\point}[1]{\par\smallskip\noindent\textbf{#1:}\xspace}

\newcommand{\empirical}[1]{{#1}}

\newcommand{\etherscanTx}[1]{\href{https://etherscan.io/tx/#1}{#1}\xspace}

\newcommand*\wrapletters[1]{\wr@pletters#1\@nil}
\def\wr@pletters#1#2\@nil{#1\allowbreak\if&#2&\else\wr@pletters#2\@nil\fi}

\usepackage{soul}

\usepackage{pdflscape}
\usepackage{rotating}
\usepackage{array}
\newcolumntype{L}{>{\centering\arraybackslash}m{3cm}}

\newtheorem{theorem}{Theorem}
\theoremstyle{definition}
\newtheorem{definition}{Definition}[section]
\hypersetup{
    colorlinks=true,
    linkcolor=blue,
    filecolor=magenta,
    urlcolor=blue,
    pdftitle={AAMM: Mitigating Frontrunning, Transaction Reordering and Consensus Instability in Decentralized Exchanges},
    pdfpagemode=FullScreen,
}

\begin{document}

\newcommand{\amm}{AMM\xspace}
\newcommand{\amms}{AMMs\xspace}
\newcommand{\aamm}{A$^2$MM\xspace}
\newcommand{\aamms}{A$^2$MMs\xspace}

\newcommand{\NumOffchainOverhead}{\empirical{$23.7$}\xspace}
\newcommand{\NumOffchainBackrunningOverhead}{\empirical{$27$}\xspace}
\newcommand{\NumOnchainBackrunningOverhead}{\empirical{$6$}\xspace}
\newcommand{\PercentageOnchainBackrunningOverhead}{\empirical{$10.1\%$}\xspace}
\newcommand{\PercentageOnchainFrontrunningOverhead}{\empirical{$4.1\%$}\xspace}
\newcommand{\AverageBytesNetworkOverhead}{\empirical{$5.83 \pm 6.57$}\xspace}
\newcommand{\AverageBytesNetworkOverheadPercentage}{\empirical{$13.8\%$}\xspace}

\newcommand{\AAMMDecreaseGasSuccessfulBackrunArb}{\empirical{$0.67\times$}\xspace}
\newcommand{\AAMMDecreaseGasFailedBackrunArb}{\empirical{$0.67\times$}\xspace}
\newcommand{\AAMMAverageProfitUSDPerSwap}{\empirical{$38.0$}\xspace}

\newcommand{\AAMMArbRouteNum}{\empirical{$460,349$}\xspace}
\newcommand{\AAMMArbRouteRatio}{\empirical{$81.87\%$}\xspace}
\newcommand{\AAMMArbRouteRevenueInETH}{\empirical{$10,675$}\xspace}
\newcommand{\AAMMArbRouteRevenueInUSD}{\empirical{$21,350,565$}\xspace}

\newcommand{\BlockSpaceReduction}{\empirical{$32.8\%$}\xspace}

\newcommand{\GasRatioAAMMSwap}{\empirical{6.97\%}\xspace}
\newcommand{\GasRatioAAMMRouting}{\empirical{17.80\%}\xspace}
\newcommand{\GasRatioAAMMRoutingArbitrage}{\empirical{60.22\%}\xspace}
\newcommand{\GasRatioAAMMArbitrage}{\empirical{42.42\%}\xspace}
\newcommand{\GasRatioSwapAndArbitrageMoreThanAAMM}{\empirical{$1.7\times$}\xspace}
\newcommand{\GasRatioAAMM}{\empirical{$1.3\times$}\xspace}

\newcommand{\ConcreteExectionStartBlock}{\empirical{$10794261$}\xspace}
\newcommand{\ConcreteExectionEndBlock}{\empirical{$12000000$}\xspace}
\newcommand{\ConcreteExectionStartDay}{\empirical{$4th$~September,~$2020$}\xspace}
\newcommand{\ConcreteExectionEndDay}{\empirical{$8th$~March,~$2021$}\xspace}
\newcommand{\ConcreteExectionDays}{\empirical{$185$}\xspace}

\newcommand{\AverageTransactionFeeReduction}{\empirical{$90\%$}\xspace}

\def\thetitle{\huge 
\aamm:\\Mitigating Frontrunning, Transaction Reordering and Consensus Instability in Decentralized Exchanges}
\title{\thetitle

}

\author{
\IEEEauthorblockN{
Liyi Zhou, 
Kaihua Qin,
Arthur Gervais
}

\IEEEauthorblockA{
Imperial College London, United Kingdom\\
}
}

\maketitle

\begin{abstract}
The asset trading volume on blockchain-based exchanges (DEX) increased substantially since the advent of Automated Market Makers (AMM). Yet, \amms and their forks compete on the same blockchain, incurring unnecessary network and block-space overhead, by attracting sandwich attackers and arbitrage competitions. Moreover, conceptually speaking, a blockchain is one database, and we find little reason to partition this database into multiple competing exchanges, which then necessarily require price synchronization through arbitrage.

This paper shows that DEX arbitrage and trade routing among similar \amms can be performed efficiently and atomically on-chain within smart contracts. These insights lead us to create a new AMM design, an Automated Arbitrage Market Maker, short \aamm DEX. \aamm aims to unite multiple \amms to reduce overheads, costs and increase blockchain security. With respect to Miner Extractable Value (MEV), \aamm serves as a decentralized design for users to atomically collect MEV, mitigating the dangers of centralized MEV relay services.

We show that \aamm offers essential security benefits. First, \aamm strengthens the blockchain consensus security by mitigating the competitive exploitation of MEV, therefore reducing the risks of consensus forks. \aamm reduces the network layer overhead of competitive transactions, improves network propagation, leading to less stale blocks and better blockchain security. Through trade routing, \aamm reduces the predatory risks of sandwich attacks by taking advantage of the minimum profitable victim input. \aamm also offers financial benefits to traders. Failed swap transactions from competitive trading occupy valuable block space, implying an upward pressure on transaction fees. Our evaluations shows that \aamm frees up \BlockSpaceReduction block-space of \amm-related transactions. In expectation, \aamm's revenue allows to reduce swap fees by~\AverageTransactionFeeReduction.

We hope that our work engenders further innovation in the space of efficient and censorship-resilient exchanges, which by design democratizes MEV and \emph{let the people trade}.
\end{abstract}

\newcommand{\Addliquidity}{{\color{purple}\small\normalfont\text{AddLiquidity}}}
\newcommand{\addliquidity}{{\color{purple}\small\normalfont\text{addLiquidity}}}

\newcommand{\Removeliquidity}{{\color{purple}\small\normalfont\text{RemoveLiquidity}}}
\newcommand{\removeliquidity}{{\color{purple}\small\normalfont\text{removeLiquidity}}}

\newcommand{\TransactXY}{{\color{purple}\small\normalfont\text{Swap$X$to$Y$}}\xspace}
\newcommand{\transactXY}{{\color{purple}\small\normalfont\text{swap$X$to$Y$}}\xspace}
\newcommand{\TransactYX}{{\color{purple}\small\normalfont\text{Swap$Y$to$X$}}\xspace}
\newcommand{\transactYX}{{\color{purple}\small\normalfont\text{swap$Y$to$X$}}\xspace}

\newcommand{\RouteXY}{{\color{purple}\small\normalfont\text{Route$X$to$Y$}}\xspace}
\newcommand{\RouteYX}{{\color{purple}\small\normalfont\text{Route$Y$to$X$}}\xspace}
\newcommand{\ArbitrageForY}{{\color{purple}\small\normalfont\text{ArbitrageFor$Y$}}\xspace}
\newcommand{\ArbitrageForX}{{\color{purple}\small\normalfont\text{ArbitrageFor$X$}}\xspace}

\newtheorem{property}{Property}

\thispagestyle{plain}
\pagestyle{plain}

\section{Introduction}\label{sec:intro}
Permissionless blockchains have portrayed their full strength when mediating financial assets among parties within censorship-resilient on-chain exchanges. One of the most popular exchange models is the Automated Market Maker~\cite{hertzog2017bancor}, where a smart contract autonomously adjusts the price for supply and demand upon incoming trading requests.

In a perfect world, different financial exchanges would all offer the same price for the same asset at the same time --- i.e., the exchanges should be perfectly synchronized. In reality, however, competing exchanges must necessarily synchronize their asset prices. High-frequency arbitrage bots are known to conduct transaction fee bidding contests, both on the blockchain's P2P network and on the consensus layer~\cite{daian2019flash, zhou2021just,zhou2020high}. Transaction fee bidding obstructs the available bandwidth on the blockchain's P2P network~\cite{daian2019flash}, therefore hinders information propagation and hence negatively influences blockchain security~\cite{gervais2015tampering}. MEV was also shown to incentivize miners to fork the chain. For example, a small rational miner with only $5$\% hashrate, will fork the Ethereum blockchain given an MEV opportunity yielding $4\times$ the block reward~\cite{zhou2021just}.

\begin{figure}[t!]
\centering
    \includegraphics[width=\columnwidth]{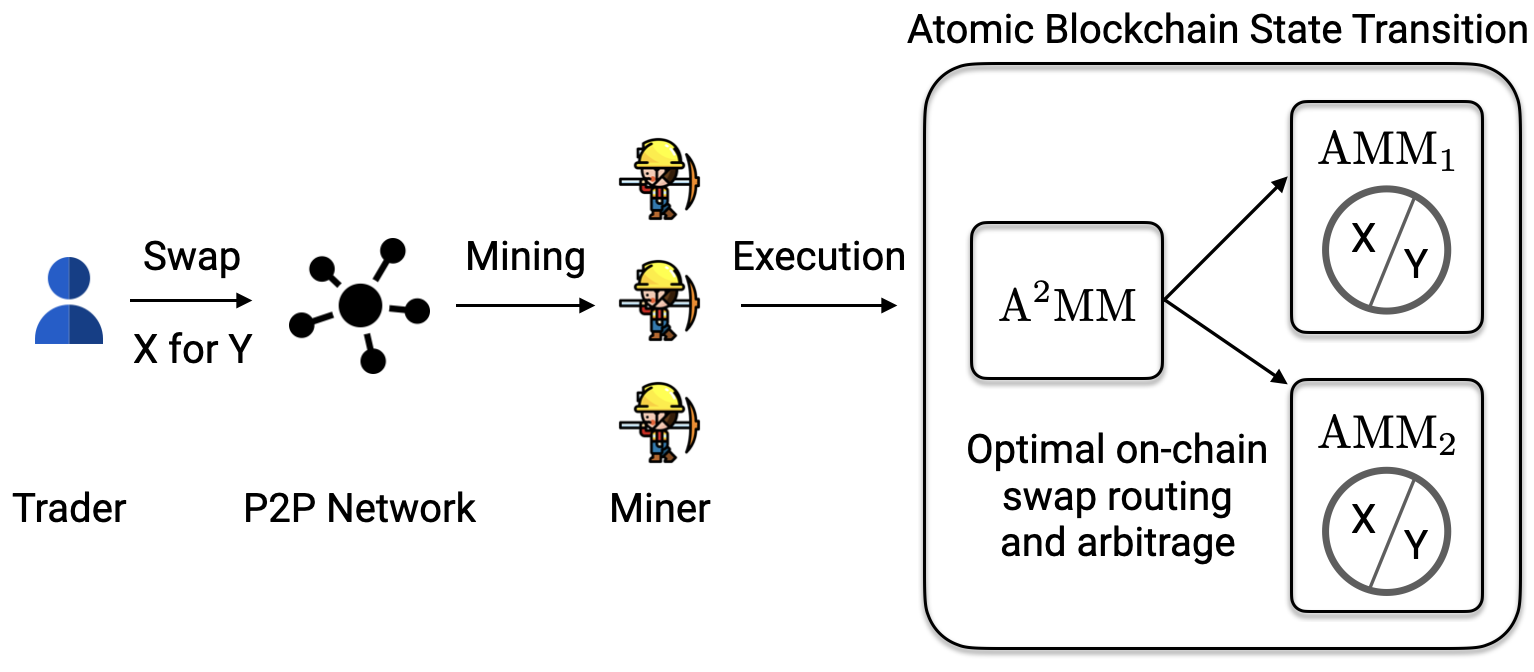}
    \caption{\aamm design, which peers with two \amms using their liquidity pools. When \aamm receives a swap transaction for a market with the assets X and Y, \aamm atomically performs optimal routing and arbitrage among the considered \amm, minimizing subsequent arbitrage transactions.}
    \label{fig:a2mm}
\end{figure}

This paper proposes a new type of \amm design, so-called Automated Arbitrage Market Maker, or \aamm, which by design performs optimal trade routing and best-effort two-point arbitrage (cf.\ Figure~\ref{fig:a2mm}) among peered \amms. \aamm offers various security benefits for the underlying blockchain. First, \aamm atomically extracts two-point arbitrage MEV from the peered \amms, which would otherwise deteriorate the blockchain's security~\cite{zhou2021just}. Second, through swap routing, \aamm reduces the risks of sandwich attacks due to the minimum profitable victim input (MVI)~\cite{zhou2020high}. Third, \aamm deters competitive network layer bidding, freeing the available blockchain network layer bandwidth, reducing the stale block rate and ultimately strengthening blockchain security~\cite{gervais2016security}.

\aamm also offers financial benefits to its users. Routing grants traders better asset prices, and arbitrage can yield positive income. Contrary to centralized MEV relayer, which auctions off MEV extraction\footnote{e.g., \url{https://github.com/flashbots/mev-relay-js}}, \ul{\emph{A$^2$MM is a decentralized and trustless design allowing users to atomically benefit from MEV, while mitigating its negative consequences}}. One drawback of \aamm is that the added smart contract logic necessarily increases the transaction fees for swaps. Our evaluation of \aamm with two \amms (cf.\ Figure~\ref{fig:a2mm}), however, shows that \aamm's routing and arbitrage in expectation allow to reduce transaction fees by~\AverageTransactionFeeReduction compared to a standard \amm swap.

\subsection*{One Blockchain --- One \amm}
The \amm design space is without a doubt considerable and multi-dimensional~\cite{xu2021sok}. Related works have for example explored the various implications of differing \amm pricing formulas~\cite{uniswap, balancerexchange, egorov2019stableswap}. Orthogonal to the pricing formula design space, we would like to propose an intuition of how multiple \amms on the same blockchain can be positively united.

Conceptually speaking, a blockchain is a distributed database, where each blockchain node aims to maintain the same view as the remaining network. If we compare a blockchain to a centralized exchange, which must also maintain its proprietary non-distributed database, then there is little reason to split such a database into multiple competing partitions, which necessarily require synchronization through price arbitrage. We therefore observe the following: \emph{(i)} multiple DEXes dilute the financial liquidity in each DEX and result in less attractive asset pricing. \emph{(ii)} multiple DEXes must synchronize through arbitrage, which causes overhead on the blockchain database and the network layer. \ul{\emph{From a security and financial perspective, it therefore appears to be strictly disadvantageous to deploy multiple DEXes on the same blockchain.}}\footnote{There may be social, competitive, and egocentric reasons for the deployment of competing DEXes (e.g., to sell a token), which we, however, do not further investigate in this work.} In this work, we take the stance that a blockchain should ideally only operate one \amm smart contract, to increase the financial efficiency, reduce network layer and block-space overhead, and consequently increase blockchain throughput as well as security.

\begin{figure}[htb!]
    \centering
    \includegraphics[width=\columnwidth]{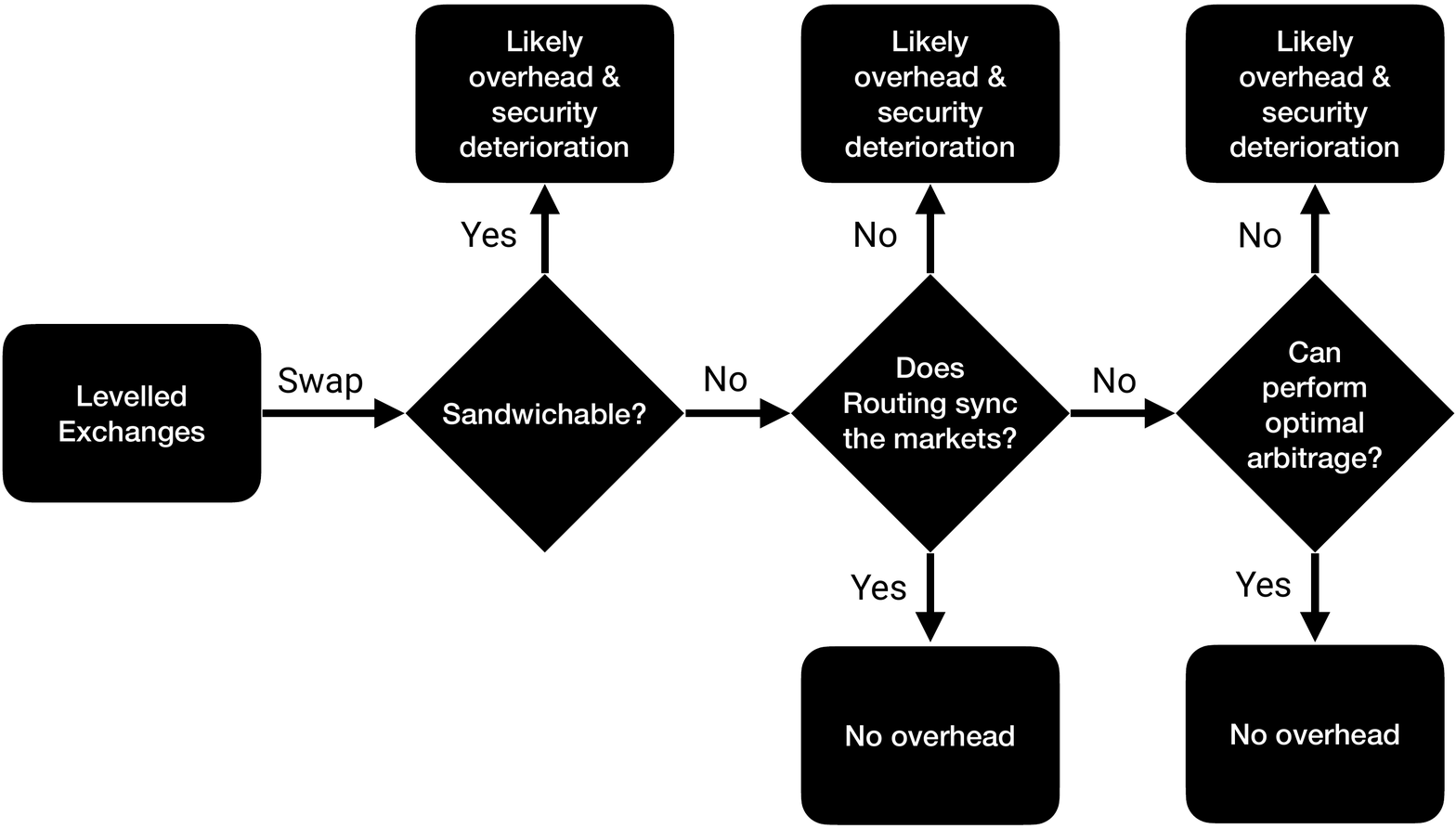}
    \caption{Decision tree whether a swap would introduce overhead in terms of the P2P network layer or block space.}
    \label{fig:decision-tree}
\end{figure}

We motivate our stance through an \amm swap decision tree in Figure~\ref{fig:decision-tree}. The decision tree departs from a possible \amm state where multiple exchanges on an identical blockchain have the same price for the same asset markets, i.e., the exchange prices are levelled. Then, a user performs a swap on one of the $i$ \amm exchanges, which would necessarily depart the \amms state from their price equilibrium. Given optimal sandwich attack parameters~\cite{zhou2020high}, we can determine whether the swap can be attacked through a sandwich attack. If the swap is not sandwichable, we find whether the AMMs can reach price equilibrium through optimal asset routing. If routing alone cannot level the prices on the different \amm markets, we can resort to arbitrage. If arbitrage can be performed atomically within the swap, we do not anticipate further overhead from the incoming swap, hence securing the blockchain from possible MEV extraction. In all other cases, there exist the likely introduction of overhead deteriorating the blockchain security, reducing its throughput and increasing transaction fees.

We summarize our main contributions in the following.

\begin{description}
\item[\aamm Design:] We provide a new AMM design, \aamm, which atomically performs optimal swap routing as well as efficient arbitrage among existing \amm liquidity pools, if deemed profitable by the \aamm smart contract. Our design does not change the simple usability aspects of existing \amms, yet allows any user to profit from atomic routing and arbitrage.

\item[\aamm Strengthens Blockchain Security:] MEV is a design problem threatening blockchain security \cite{qin2021quantifying,zhou2021just}. \aamm mitigates two MEV sources, namely two-point arbitrages and sandwich attacks. We show that adopting an \aamm design reduces both block-space and network layer overhead caused by MEV bots, therefore, strengthening the blockchain consensus by reducing the stale block rate. We find that $88.80\%$ of the back-running arbitrage transactions are accompanied by what we call \emph{back-run flooding} (BRF), an observed denial of service practice on the blockchain P2P layer.





\item[Evaluation:] We implement and evaluate \aamm as shown in Figure~\ref{fig:a2mm}, while synchronizing with two \amms (i.e., Uni- and Sushiswap). By replaying past blockchain data, through routing and arbitrage revenue, on average, we find that \aamm reduces the consumed transaction fees of a standard \amm swap by~\AverageTransactionFeeReduction. Moreover, in expectation, \aamm reduces the consumed block-space by~\BlockSpaceReduction.
\end{description}


The remainder of the paper is organized as follows. Section~\ref{sec:background} provides a background, while Section~\ref{sec:aamm} introduces our system and threat model and outlines the \aamm design. Section~\ref{sec:evaluation} presents our evaluation and empirical results. We highlight \aamm's security implications in Section~\ref{sec:security-implications} and shed light on the cost when \aamm peers with more than two \amms in Section~\ref{sec:cost_multiple}. We summarize related works in Section~\ref{sec:related-work}, provide a discussion in Section~\ref{sec:discussion} and conclude the paper in Section~\ref{sec:conclusion}.

\section{Decentralized Finance (DeFi)}\label{sec:background}
Since the inception of permissionless blockchains with Bitcoin in 2008~\cite{bitcoin}, it became apparent that their most well-suited use case is the transfer or trade of financial assets without trusted intermediaries~\cite{wust2018you}. A blockchain is considered permissionless when entities can join and leave the network at any point in time. Users authorize transactions through a public key signature and a subsequent broadcast on the blockchain P2P network. The formatted public key corresponds to an address that a user can receive assets at. Miners accumulate transactions and solve a proof of work (PoW) puzzle to append a block to the blockchain (various alternatives to PoW emerged, such as PoS~\cite{saleh2021blockchain,bano2019sok}). Miners are financially rewarded for performing work for the network through block rewards and transaction fees. A third miner reward source, which is gaining traction~\cite{qin2021quantifying}, is the extraction of Miner Extractable Value. While Bitcoin supports basic smart contracts through a stack-based programming language, the support for loops and higher-level languages (such as Solidity) have gained widespread adoption. For a more thorough blockchain background, we refer the reader to several helpful SoKs~\cite{bonneau2015sok,atzei2017survey,bano2017consensus}.

Smart contracts provide the building blocks for an ecosystem of decentralized finance~\cite{schar2020decentralized}, where users can interact with lending pools, AMM exchanges, stablecoins, derivatives, asset management platforms etc. At the time of writing, DeFi has grown to an accumulative locked value of over $60B$ USD. For a more thorough background on DeFi, we refer the interested reader to an SoK~\cite{schar2020decentralized}. We proceed to separate the background into a finance- and security-related overview.

\subsection{Finance Background}

\point{Market Maker} Market makers (MM) help the market (i.e.,\ the traders buying and selling assets) having access to sufficient liquidity (i.e.,\ monetary asset amounts) for buy/sell orders to match at the ask/bid price. Traditionally, market makers are incentivized to operate as they can profit from the spread (i.e.,\ the difference) between the bid and ask prices.

\point{Automated Market Maker Exchanges} AMMs govern through smart contracts a pool of assets, where a pricing formula defines the asset purchase and sell price. Several \amm pricing formulas are proposed in the literature; the most popular form is the constant product AMM~\cite{hertzog2017bancor}. While Bancor introduced the \amm concept, at the time of writing, Uniswap~\cite{uniswap} is the most prominent \amm with a daily trading volume of over $944M$ USD and $5.06B$ USD of the total supplied liquidity among $31,202$ different asset pairs\footnote{\url{https://info.uniswap.org/}}. One of the better-known forks of Uniswap is Sushiswap~\cite{sushiswap}.

\point{Arbitrage}
The process of selling/buying an asset in one market while concurrently buying/selling in another market at a different price is known as arbitrage. Arbitrage encourages economic stability and is generally regarded as benign. DeFi traders/miners track new blockchain state changes and conduct arbitrage if the anticipated revenue from synchronizing two markets exceeds the expected transaction costs. To perform arbitrage, a trader may operate on the previous block state or on the state of the pool of unconfirmed transactions (i.e.\ the mempool)~\cite{qin2021quantifying}.

\point{Slippage}
The adjustment in the price of an asset during a transaction is known as price slippage. Expected price slippage is the anticipated rise or decrease in price depending on the amount to be exchanged and the available liquidity~\cite{zhou2020high}. The expected slippage increases as trading volume increases. Unexpected price slippage is the rise or decrease in price during the interim time between creating a transaction and its execution. The sum of the expected and unexpected slippage represents the price impact of a trade.

\point{Swap Routing Aggregators}
An exchange aggregator is a service to aggregate liquidity from multiple exchanges. Aggregators may split a single trade into numerous smaller transactions to receive the best overall trade price. The sub-trades are then routed to various exchanges to provide the best exchange price and minimize the trading slippage. In March 2021, the three most significant off-chain aggregators (1inch, Mocha, and Paraswap) amassed a total monthly volume of~$11.49B$~USD~\footnote{\url{https://www.theblockcrypto.com/data/decentralized-finance/dex-non-custodial/dex-aggregator-trade-volume}}. Off-chain aggregators are not guaranteed to yield optimal execution parameters due to the unexpected slippage. To the best of our knowledge, off-chain aggregators also do not perform arbitrage.

\point{Flash Loans}
Atomic blockchain transactions may execute several actions in a rigorous sequence. If a single transaction fails in one of its execution steps, the entire transaction fails atomically and does not alter the blockchain state. This atomicity property enables a novel financial product, flash loans. Flash loans are loans drawn from a smart contract pool of assets and are only valid within one atomic transaction. The flash loan must be paid back by the end of the transaction; otherwise, the loan fails. When a flash loan fails, the blockchain state is not modified, corresponding to a state where the loan was never granted to the borrower. Because lenders bear no risks by the borrowers defaulting on the loan, flash loans quickly grow to billions of USD~\cite{qin2020attacking}.

\subsection{Security Background}

\point{Front- and Back-running} is the process by which an adversary observes a victim's pending transaction on the network layer and then acts upon this information by placing trades before or after the victim's transaction. While custodian and centralized financial services are known to be under the supervision of regulatory bodies~\cite{finma-ban}, DeFi (and blockchain) are not yet thoroughly regulated. Previous studies have observed front-running bidding wars between DEX arbitrage bots~\cite{daian2019flash,qin2021quantifying}. Transaction fee bidding causes on-chain congestion and introduces network layer overhead, which necessarily increases the stale block rate and hence weakens the consensus security of the underlying blockchain~\cite{gervais2016security}.

\point{Sandwich Attack}
A sandwich attack is a predatory trading strategy, which exploits a pending, not yet executed trade~\cite{zhou2020high}. Suppose an asset's price is expected to rise/fall due to a pending trade. In that case, a malicious front-runner can buy/sell the asset before the victim transaction executes and then close its position by selling/buying the same asset after the victim transaction is confirmed. Because \amms provide complete transparency about the exchange's state and the pricing formula, sandwich attackers can derive the optimal attack parameters. Previous works investigate AMM-specific mitigations and find that sandwich attacks are not profitable if the victim's input amount remains below a safe, market-state-specific threshold.

\point{Miner Extractable Value}
Miners retain the authority to decide on the transaction order of their mined blocks. Miners observe pending transactions on the network layer and may maximize their revenue through the optimal transaction order. For instance, the miners can perform front-running or sandwich attacks~\cite{zhou2020high}. The term ``Miner Extractable Value'' or MEV~\cite{daian2019flash}, refers to the entire potential revenue that miners can extract through transaction order manipulation. Related work quantified that at least~$28.8M$ USD in profit was extracted over two years following the~$1$st December~$2018$~\cite{qin2021quantifying}. Because non-MEV miners order by default transactions in descending transaction fee (gas price) amount~\cite{zhou2020high}, a non-mining trader can also capture MEV by adjusting their transaction fees. Related work, for example, shows how trading bots engage in competitive transaction fee bidding contests~\cite{qin2021quantifying}.

\point{Stale Block Rate}
Previous works have extensively shown that blockchain forks increase the stale block rate and deteriorate the consensus security by increasing the risks of double-spending and selfish mining~\cite{gervais2016security,eyal2014majority,bonneau2016buy}. Zhou~\etal~\cite{zhou2021just} quantified an MEV threshold at which MEV-aware miners are incentivized to fork the blockchain. For instance, on Ethereum, a miner with a hash rate of $10\%$ would fork the blockchain if an MEV opportunity exceeds $4$× the block reward. Because arbitrage is one of the prime sources of MEV, it is therefore of utmost importance to minimize the need for arbitrage to increase blockchain security.

\subsection{\amm Arbitrage}
\begin{figure}[htb!]
    \centering
    \includegraphics[width=\columnwidth]{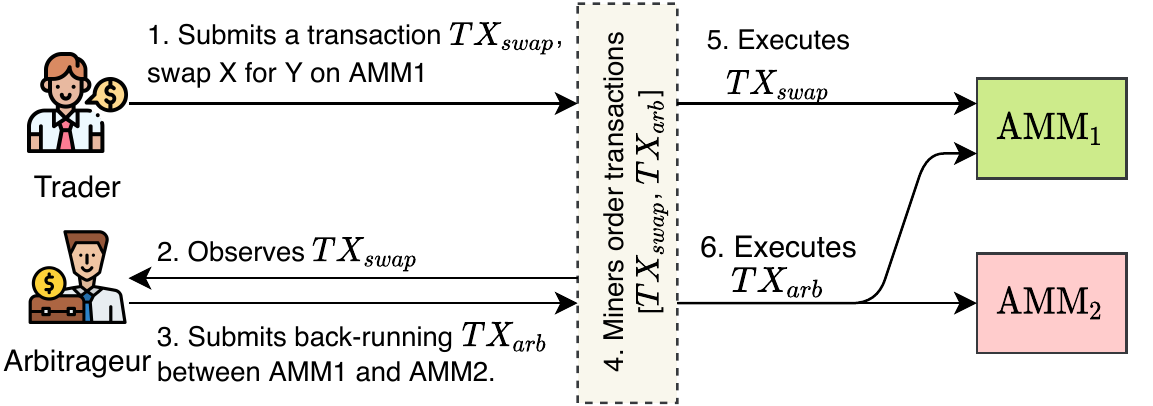}
    \caption{Overview of the back-running arbitrage process in \amm exchanges. The liquidity taker initiates a swap on AMM1 by broadcasting its transaction ($TX_{swap}$) on the P2P network. An arbitrageur listens on the public P2P network and observes $TX_{swap}$. The arbitrageur then issues a back-running arbitrage transaction ($TX_{arb}$), if $TX_{swap}$ creates a profitable arbitrage opportunity between AMM1 and AMM2. Note that miners can collude with arbitrageurs to extract profits without failure risks.}
    \label{fig:amm}
\end{figure}

We outline the traditional \amm design coupled with the necessary third-party arbitrageurs in Figure~\ref{fig:amm}. This \amm design necessarily requires at least two separate transactions, $TX_{swap}$ and $TX_{arb}$ to synchronize the prices on AMM1 and AMM2 after the swap. Moreover, because $TX_{swap}$ and $TX_{arb}$ are non-atomic (i.e.,\ do not necessarily execute in succession), multiple arbitrageurs, as well as miners, are likely to compete over benefiting from the created arbitrage opportunity~\cite{daian2019flash}.

\subsection{Off-chain AMM routing}
\begin{figure}[htb!]
    \centering
    \includegraphics[width=\columnwidth]{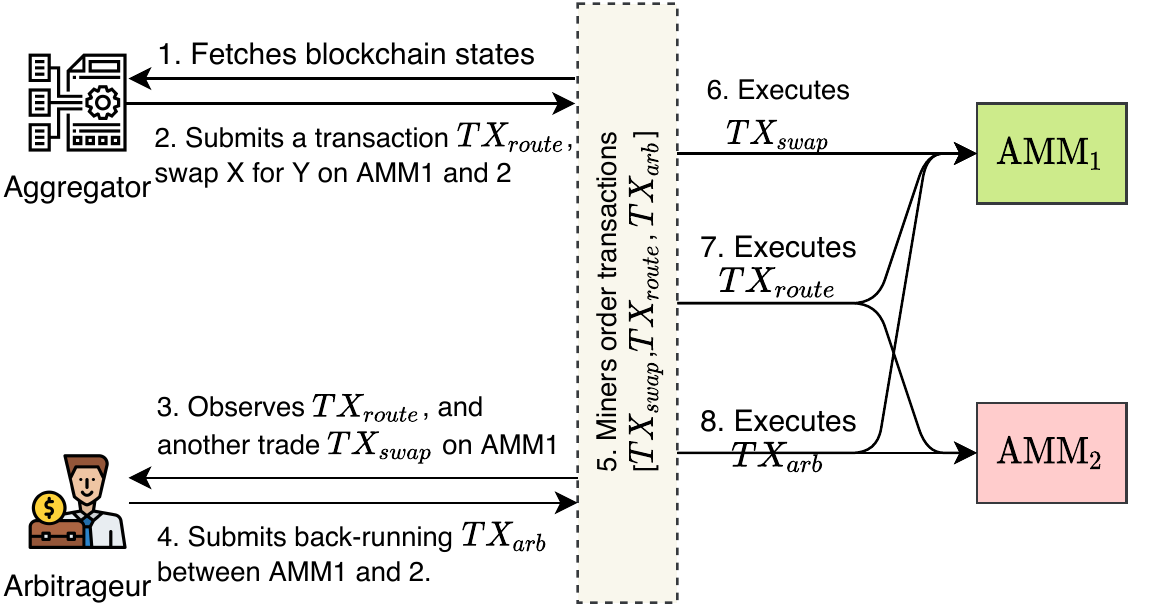}
    \caption{Off-chain routing aggregator. Upon (0) incoming swap, the routing service calculates (1) optimal paths and trading parameters given a blockchain state. The routing service (2) issues the transaction ($tx_{route}$) on the P2P network. An arbitrageur (3) observes $tx_{route}$, and another swap ($tx_{swap}$). The arbitrageur then (4) performs a back-running arbitrage ($tx_{arb}$). Because the transaction order execution is not guaranteed off-chain routing paths and parameters may be suboptimal.
    }
    \label{fig:agg}
\end{figure}

Off-chain swap routing services calculate the best routing path and parameters based on their local blockchain state (cf.\ Figure~\ref{fig:agg}). Off-chain routing avoids complex smart contract operations, thereby minimizing transaction costs. However, off-chain routing paths and parameters are not necessarily optimal during execution, because the blockchain state might change intermittently between route generation and execution. To mitigate this problem, aggregators (such as 1inch) cooperate with miners on privately mined transactions, which arguably renders DeFi more centralized~\cite{1inch-frontrunning,qin2021quantifying}. Moreover, the goal of routing is to only optimize the liquidity takers' swap transactions, while forgoing possible arbitrage opportunities between other AMM exchanges. 

\section{\aamm}\label{sec:aamm}
We proceed to introduce the \aamm design, system, threat, and state transition model by adding arbitrage actions on top of a known \amm model~\cite{zhou2020high}. Note that we only study two-point arbitrages, and we, therefore, focus on markets with two assets in the following. 

\subsection{System Model}
We consider a blockchain P2P network, where traders interact with AMMs by signing transactions with their respective private keys. For example, traders can exchange cryptocurrency assets, deposit/withdraw assets to/from different exchange pools, perform arbitrages, etc. Traders can freely adjust the parameters of these transactions, such as the transaction fees (e.g., gas price), slippage limit, expiration time, etc. A trader then broadcasts a transaction on the asynchronous blockchain peer-to-peer (P2P) network~\cite{decker2013information,kim2018measuring,kifferunder}, or may privately send transactions to miners~\cite{zhou2020high}. The transaction propagation typically utilizes gossip or publish-subscribe mechanisms, and nodes (including miners) may have different views of the pending (i.e.,\ unconfirmed) transactions stored in the mempool. By default, miners order transactions according to the paid transaction fees but were also shown to adhere to private ordering policies~\cite{qin2021quantifying}.

\subsection{Threat Model}
We do not constrain the mining behavior of the miners but assume that no miner can accumulate more than 33\% of the total hash-rate~\cite{eyal2014majority}. Miners can manipulate transaction ordering by transparently ignoring the default transaction ordering rules (i.e., highest-priced transactions first) or by attempting to hide private agreements by pretending to participate in transaction fee bidding contests~\cite{qin2021quantifying}. We assume that smart contracts are secure and free from vulnerabilities.

\subsection{\amm State Transition}
We follow the standard model for \amm exchanges~\cite{zhou2020high}. An \amm consists of mainly two types of \textbf{traders}, namely the \textbf{liquidity providers} and \textbf{liquidity takers}. A liquidity taker buys or sells an asset in exchange for another asset, using the liquidity providers' disposable assets.

\begin{tcolorbox}[title = AMM State,arc=0pt,outer arc=0pt]
\begin{definition}\label{def:state}
The state (or depth) of an AMM market with two assets $X$ and $Y$ is defined as $(x,y)\in (\mathbb{R}^{+})^2$. The sum of $x$ and $y$ correspond the total amount of assets from $X$ and $Y$ deposited by liquidity providers.
\end{definition}
\end{tcolorbox}
\vspace{1em}

Two-asset \amms typically support the following two-state transition functions for liquidity takers to convert between asset X and Y.

\begin{enumerate}
    \item \TransactXY: A liquidity taker can trade $\delta_x$ of asset $X$, increasing the available liquidity of asset $X$, in exchange for $\delta_y$ of asset $Y$, decreasing the available liquidity of asset $Y$ (cf.\ Equation~\ref{eq:AMM_transact_y_for_x}).
    \begin{equation}\label{eq:AMM_transact_y_for_x}
    \begin{split}
    (x,y) &\xrightarrow[\delta_x \in \mathbb{R}^{+}]{\TransactXY(\delta_x, \delta_y)} (x + \delta_x,\ y - \delta_y)\\
    \text{where:\quad} &\delta_y \leq \Delta_y = p_{X \rightarrow Y}(x, y, \delta_x) \cdot \delta_x
    \end{split}
    \end{equation}
    \item \TransactYX: The mirroring action for \TransactXY.
\end{enumerate}

Note that both \TransactXY and \TransactYX use a pricing function $p(\cdot)$ to determine the maximum amount of asset $Y$ the taker can receive. Each AMM exchange may choose a custom pricing function $p(\cdot)$ for governing the asset exchange. A liquidity taker can exchange $\delta_x$ amount of asset $X$ for up to $\Delta_y$ amount of asset $Y$, while choosing to withdraw fewer assets $Y$ voluntarily.

\begin{tcolorbox}[title=Pricing Formula,arc=0pt,outer arc=0pt]
\begin{definition}\label{def:pricing}
A pricing formula is a differentiable function $p_{X \rightarrow Y}(x, y, \delta_x) : (\mathbb{R}^{+})^3 \mapsto \mathbb{R}^{+}$, which maps the \amm state ($x, y$) and input amount ($\delta_x$) of asset $X$ to the best exchange rate the \amm offers.
\end{definition}
\end{tcolorbox}
\vspace{1em}

\point{Assumptions}
We assume that the \amms we consider abide by the following properties.

\begin{tcolorbox}[title=Liquidity Sensitivity,arc=0pt,outer arc=0pt]
\begin{property}\label{prop:liquidity_sensitivity} Given an AMM state $(x, y)$, the price $p_{X\rightarrow Y}(x, y, \delta_x)$ decreases as the trade size ($\delta_x$) increases. Similarly, the price $p_{Y\rightarrow X}(x, y, \delta_y)$ decreases as the trade size ($\delta_y$) increases.
\end{property}
\end{tcolorbox}
\vspace{1em}

Liquidity sensitivity (cf.\ Property~\ref{prop:liquidity_sensitivity}) enables the underlying \amm market to adjust autonomously the price based on the trading volume and direction. The more asset $X$ a liquidity taker purchases from an AMM, the more scarce $X$ becomes in the liquidity pool, and therefore the price of $X$ increases relative to $Y$ (and vice versa). Property~\ref{prop:liquidity_sensitivity} implies that the pricing functions are monotonically decreasing as the trade volume increases.\\

\begin{tcolorbox}[title=Path Independence,arc=0pt,outer arc=0pt]
\begin{property}\label{prop:path_independence} 
Given an inital market state $(x, y)$, the following sub-properties holds:
\begin{enumerate}
    \item Two consecutive \TransactXY transactions, respectively swapping $\delta_x^1, \delta_x^2$ asset $X$ to $\delta_y^1, \delta_y^2$ asset $Y$, are equivalent to one \TransactXY transactions, swapping $\delta_x^1 + \delta_x^2$ asset $X$ to $\delta_y^1 + \delta_y^2$ asset $Y$.
    \item Two consecutive transactions, swapping $\delta_x^1$ asset $X$ to $\delta_y^1$ asset $Y$(\TransactXY) followed with $\delta_y^2$ asset $Y$ back to $\delta_x^2$ asset $X$(\TransactYX), where $\delta_x^1 - \delta_x^2 = \delta_x$, are equivalent to one \TransactXY transaction, swapping $\delta_x^1 - \delta_x^2$ asset $X$ for $\delta_y^1-\delta_y^2$ asset $Y$.
\end{enumerate}
\end{property}
\end{tcolorbox}
\vspace{1em}

Path Independence is a desirable \amm property because it ensures that liquidity takers have no incentive to split a trade into multiple smaller transactions on the \textbf{same} \amm market. Note that when there exist numerous appropriate \amm markets, it can still be more profitable to split a trade and perform routing (cf.\ Section~\ref{sec:design}).\\

\begin{tcolorbox}[title=Market Independence,arc=0pt,outer arc=0pt]
\begin{property}\label{prop:market_independence} 
Given two state transition actions on two different \amm markets, the execution order of these two transitions will not impact the final states of the two AMMs.
\end{property}
\end{tcolorbox}
\vspace{1em}

Property~\ref{prop:path_independence} and~\ref{prop:market_independence} are applied in the following to compress routing and arbitrage transactions in Section~\ref{sec:compress}. Note that the same \amm can have multiple markets trading the same asset pairs ($X,Y$), but we assume these markets to have different states. A state change on one market will hence not affect the state or price of another market.  

\subsection{\aamm Design}\label{sec:design}
\begin{figure}[tb!]
    \centering
    \includegraphics[width=\columnwidth]{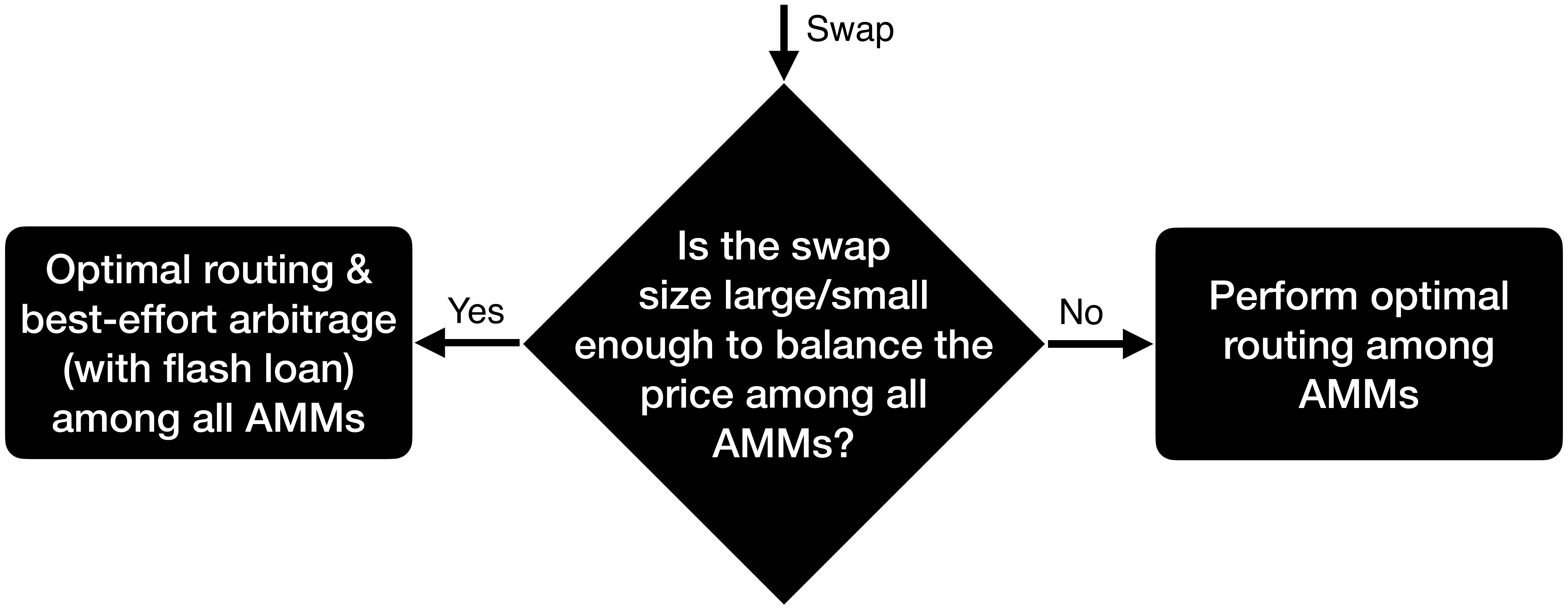}
    \caption{Decision tree of the \aamm exchange logic. \aamm encourages first optimal routing, and if routing alone isn't sufficient, best-effort arbitrage may synchronize the markets.}
    \label{fig:a2mm-decisiontree}
\end{figure}

In the following section we describe the proposed \aamm design (cf.\ Figure~\ref{fig:a2mm}). On a high-level, \aamm minimizes the arbitrage opportunities between itself and other \amm exchanges, after any \aamm state transition (e.g., a \emph{swap}, \emph{add} or \emph{remove} liquidity). Upon an incoming swap, the \aamm first checks if the swap amount is sufficiently large to synchronize the prices on the considered AMMs, and then performs optimal routing (cf.\ Figure~\ref{fig:a2mm-decisiontree}). Otherwise, \aamm performs optimal routing and best-effort arbitrage among the considered AMMs. A flash loan can be requested if the swap's trader holds an insufficient balance for arbitrage~\cite{qin2020attacking}.

We assume that an \aamm market ($X,Y$) is synchronizing with $N$ other \amm $X,Y$ markets. The state of the $i$th ($1 \leq i \leq N$) \amm $X,Y$ market is denoted as $(x_i, y_i)$. We formally introduce the state transitions for arbitrage and routing.

\begin{enumerate}
    \item \ArbitrageForX: An arbitrageur can perform an arbitrage between two or multiple \amm exchanges. Given two AMMs with states $(x_i, y_i)$ and $(x_j, y_j)$ respectively. The trader initiates the arbitrage by swapping $\delta_x$ of $X$ for $\delta_y$ amount of $Y$ on \amm $i$. The arbitrageur then reverses the trade by exchanging $\delta_y$ amount of  $Y$ on \amm $j$ for $\delta_x'$ amount of $X$. If the arbitrage is successful, the arbitrageur gains $\delta_x' - \delta_x$ amounts of $X$ (cf.\ Equation~\ref{eq:AMM_arbitrage}).
    \begin{equation}\label{eq:AMM_arbitrage}
    \begin{split}
    &(x_i,y_i,x_j,y_j) \xrightarrow[\delta_x \in \mathbb{R}^{+}]{\ArbitrageForX(\delta_x, i, j)} \\
    & \qquad\qquad (x_i + \delta_x, y_i - \delta_y, x_j - \delta_x', y_j + \delta_y) \\ 
    &\Longleftrightarrow  \\
    &(x_i,y_i) \xrightarrow[\delta_x \in \mathbb{R}^{+}]{\TransactXY(\delta_x, \delta_y)} (x_i + \delta_x,\ y_i - \delta_y)\\
    &(x_j,y_j) \xrightarrow[\delta_y \in \mathbb{R}^{+}]{\TransactYX(\delta_y, \delta_x')} (x_j - \delta_x',\ y_j + \delta_y) \\
    &\text{given\ } \delta_x' - \delta_x > 0, \text{s.t.} \begin{cases}
    \delta_y \leq p_i^{X \rightarrow Y}(x_i, y_i, \delta_x) \cdot \delta_x \\
    \delta_x' \leq p_j^{Y \rightarrow X}(x_j, y_j, \delta_y) \cdot \delta_y 
    \end{cases}
    \end{split}
    \end{equation}
    \item \ArbitrageForY: The mirroring action for \ArbitrageForX.
    \item \RouteXY: When a liquidity taker swaps $\delta_x$ of asset $X$ for $\delta_y$ of asset $Y$, the taker can split its trade ($\delta_{x1},\dots,\delta_{xN}$) across $N$ \amms (cf.\ Equation~\ref{eq:AMM_routXY}).
    \item \RouteYX: The mirroring action for \RouteYX.
    
    \begin{equation}\label{eq:AMM_routXY}
    \begin{split}
    &(x_1, y_1, \dots, x_{N}, y_{N}) \xrightarrow[\delta_{xi} \in \mathbb{R}^{+},\quad \forall 1 \leq i \leq N]{\RouteXY(\delta_{x})} \\
    &\qquad\qquad(x_1 + \delta_{x1}, y_1-\delta_{y_1}, \dots, y_N - \delta_{y_N}) \\
    &\Longleftrightarrow \\
    &(x_i,y_i) \xrightarrow[\delta_{xi} \in \mathbb{R}^{+}, \forall 1 \leq i \leq N]{\TransactXY(\delta_{xi}, \delta_{yi})} (x_i + \delta_{xi},\ y_i - \delta_{yi})  \\
    &\qquad\text{s.t.\ } \begin{cases}
    \delta_x = \sum_{i=1}^N \delta_{xi}, \quad \delta_y = \sum_{i=1}^N \delta_{yi} \\
    \delta_{yi} \leq p_i^{X \rightarrow Y}(x_i, y_i, \delta_{xi}) \cdot \delta_{xi}
    \end{cases}
    \end{split}
    \end{equation}
\end{enumerate}

\subsection{Optimal On-Chain Swap Routing}\label{sec:optimal-on-chain-swap-routing}
\aamm performs \RouteXY to maximize the amount of asset $Y$ the liquidity taker purchases after paying $\delta_x$. We proceed to formally define the optimal routing problem among $N$ \amms.

\begin{equation}\label{eq:route_obj}
\begin{split}
& \text{maximize}  \ \delta_y = \sum_{i=1}^N \delta_{yi} \quad \text{given}     \ (x_i, y_i)\ \forall\ 1 \leq i \leq N  \\
& \text{s.t.} \  (x_1, y_1, \dots, x_{N}, y_{N}) \xrightarrow[\delta_{xi} \in \mathbb{R}^{+},\ \forall 1 \leq i \leq N]{\RouteXY(\delta_{x}), \delta_x = \sum_{i=1}^N \delta_{xi}} \\
    & \quad (x_1 + \delta_{x1}, x_1-\delta_{y_1}, \dots, x_N + \delta_{x_N}, y_N - \delta_{y_N}), \\
     & 
\end{split}
\end{equation}

\begin{tcolorbox}[arc=0pt,outer arc=0pt]
\begin{theorem}\label{theorem:route-lowest-price}
Routing optimization aims to level the asset price on multiple \amms and can be solved by greedily routing transaction volume.
\end{theorem}
\end{tcolorbox}

PROOF BY CONTRADICTION: This proof shows that the optimal routing among N \amms must greedily route transaction volume to the exchange with the best price (cf.\ Theorem~\ref{theorem:route-lowest-price}). We assume the existence of an optimal routing strategy ($S_{\text{optimal}}$) for \RouteXY, which in total routes $\delta_{x}$ amount of asset $X$ to $\delta_{y}$ amount of asset $Y$. More specifically, this optimal strategy routes $\delta_{x1}, \ldots, \delta_{xN}$ of asset $X$ to N \amms, in exchange of $\delta_{y1}, \ldots, \delta_{yN}$ of asset $Y$ ($\delta_{x} = \sum_{z=1}^N \delta_{xz}$, $\delta_{y} = \sum_{z=1}^N \delta_{yz}$). After the routing, we assume that \amm 2 still offers a better price than \amm 1, meaning that $S_{\text{optimal}}$ contradicts Theorem~\ref{theorem:route-lowest-price} and does not route all trading volume greedily to the \amm with the best price. Equation~\ref{eq:non-sync_output} shows the state transition process.

\begin{equation}\label{eq:non-sync_output}
\begin{split}
& (x_1, y_1)  \xrightarrow{\TransactXY(\delta_{x1}, \delta_{y1})} (x_1', y_1') \\
& (x_2, y_2)  \xrightarrow{\TransactXY(\delta_{x2}, \delta_{y2})} (x_2', y_2') \\
& \delta_{y} = p_{X \rightarrow Y}(x_1, y_1, \delta_{x1})\delta_{x1} \\
& \qquad  + p_{X \rightarrow Y}(x_2, y_2, \delta_{x2})\delta_{x2} + \sum_{z=3}^{N} \delta_{yz}
\end{split}
\end{equation}

To prove that $S_{\text{optimal}}$ is not the optimal routing strategy, we show that the routing can output more asset $Y$ if more trading volume is routed to \amm 2, without changing the routing strategy for \amm 3 to N. We denote this alternative strategy as $S_{\text{alter}}$, which routes $\delta_{x1}-\Delta, \delta_{x2}+\Delta$ of asset $X$ to \amms 1 and 2 respectively. \amm 2 still offers a better price for \transactXY after executing $S_{alter}$, because the additionally routed amount ($\Delta$) is arbitrarily small and $P_{X \rightarrow Y}$ is a continuous function. Equation~\ref{eq:non-sync_output2} shows the state transition process for $S_{\text{alter}}$.

\begin{equation}\label{eq:non-sync_output2}
\begin{split}
& (x_1, y_1)  \xrightarrow{\TransactXY(\delta_{x1}-\Delta, \delta_{y1}*)} (x_1*, y_1*) \\
& (x_2, y_2)  \xrightarrow{\TransactXY(\delta_{x2}+\Delta, \delta_{y2}*)} (x_2*, y_2*) \\
& \delta_y' =  p_{X \rightarrow Y}(x_1, y_1, \delta_{x1}-\Delta)\cdot(\delta_{x1}-\Delta) \\
& \qquad + p_{X \rightarrow Y}(x_2, y_2, \delta_{x2}+\Delta)\cdot(\delta_{x2}+\Delta) + \sum_{z=3}^{N} \delta_{yz} \\
& \text{where:}\ p_{X \rightarrow Y}(x_1*, y_1*, \Delta) < p_{X \rightarrow Y}(x_2*, y_2*, \Delta) \\
\end{split}
\end{equation}

Based on the liquidity sensitivity property (Property~\ref{prop:liquidity_sensitivity}), \amm 2 offers a worse price for \TransactXY after executing $S_{\text{alter}}$ compared to  executing $S_{\text{optimal}}$ (cf.\ Equation~\ref{eq:non-sync4}). This is because $S_{\text{alter}}$ routes more trade volume to \amm 2, where both strategies have the same initial state for \amm 2 $(x_2, y_2)$.

\begin{equation}\label{eq:non-sync4}
\small
p_{X \rightarrow Y}(x_2*, y_2*, \Delta) < p_{X \rightarrow Y}(x_2', y_2', \Delta)
\end{equation}

In Equation~\ref{eq:non-sync3} we derive that the amount of asset $Y$ $S_{\text{alter}}$ outputs is greater than $S_{\text{optimal}}$ using the path independence property (cf.\ Property~\ref{prop:path_independence}), which contradicts the assumption that $S_{\text{optimal}}$ is the optimal routing strategy. Theorem~\ref{theorem:route-lowest-price} is therefore proven by contradiction. $\square$

\begin{equation}\label{eq:non-sync3}
\small
\begin{split}
    & \delta_y' - \delta_y =  \ p_{X \rightarrow Y}(x_1, y_1, \delta_{x1}-\Delta)\cdot(\delta_{x1}{-}\Delta) \\
    & \quad + p_{X \rightarrow Y}(x_2, y_2, \delta_{x2}+\Delta)\cdot(\delta_{x2}{+}\Delta) \\
    & \quad - p_{X \rightarrow Y}(x_1, y_1, \delta_{x1})\cdot\delta_{x1} - p_{X \rightarrow Y}(x_2, y_2, \delta_{x2})\cdot\delta_{x2} \\
    & =\ \hcancel{p_{X \rightarrow Y}(x_1, y_1, \delta_{x1}{-}\Delta)\cdot(\delta_{x1}{-}\Delta)} + \hcancel{p_{X \rightarrow Y}(x_2, y_2, \delta_{x2})\cdot\delta_{x2}}\\
    & \quad + p_{X \rightarrow Y}(x_2', y_2', \Delta)\cdot\Delta \\
    & \quad - \hcancel{p_{X \rightarrow Y}(x_1, y_1, \delta_{x1}{-}\Delta)\cdot(\delta_{x1}{-}\Delta)} - \hcancel{p_{X \rightarrow Y}(x_2, y_2, \delta_{x2})\cdot\delta_{x2}} \\
    & \quad - p_{X \rightarrow Y}(x*_1, y*_1, \Delta)\cdot\Delta\\
    & = \ p_{X \rightarrow Y}(x_2', y_2', \Delta)\cdot\Delta - p_{X \rightarrow Y}(x_1*, y_1*, \Delta)\cdot\Delta \\
    &\geq\ p_{X \rightarrow Y}(x_2', y_2', \Delta)\cdot\Delta - p_{X \rightarrow Y}(x_2*, y_2*, \Delta)\cdot\Delta \geq 0
\end{split}
\end{equation}


\subsection{Arbitrage Profit Maximization}
In the following, we formally introduce the arbitrage profit maximization problem between $N$ \amms. An arbitrage between multiple DEXes may include multiple sub-arbitrage steps. Given an arbitrage strategy with $L$ steps, we use the superscript, such as $x^l$, $y^l$, to denote the state and parameters at a sub-step $l$, where $1 \leq l \leq L$. The objective of the arbitrageur is to maximize the revenue after executing all sub-arbitrage steps (cf.\ Equation~\ref{eq:arb_obj}). Because the solution of Equation~\ref{eq:arb_obj} depends on the implementation-specific \amm pricing formulas for the $N$ AMMs, we do not provide a general solution here. The reader, however, can find an optimal solution for two \amms with constant product pricing formula in Section~\ref{sec:evaluation_arbitrageforx} (corresponding to Uni- and Sushiswap capturing over~$73.27\%$ of the total \amm market trading volume at the time of writing\footnote{\url{https://www.theblockcrypto.com/data/decentralized-finance/dex-non-custodial/dex-volume-monthly}}, accessed March 2021).

\begin{equation}\label{eq:arb_obj}
\small
\begin{split}
\text{given:}           &\ (x_i, y_i)\ \forall\ 1 \leq i \leq N \\
\text{maximize:}        &\ \sum^{L}_{l=0} \delta'^{l}_x-\delta^{l}_x \\
\text{subject to:}      &\ \text{for each sub-step }l \text{ (in total }L\text{ steps)}: \\
                        & \begin{cases} 
                        (x^{l+1}_{i_l}, y^{l+1}_{i_l}, x^{l+1}_{j_l}, y^{l+1}_{j_l}) \\
                        \qquad = \ArbitrageForX(\delta^{l}_x, i_l, j_l) \\
                        \qquad = (x^{l}_{i_l} + \delta^{l}_x, y^{l}_{i_l} - \delta^{l}_y, x^{l}_{j_l} - \delta'^{l}_x, y^{l}_{j^l} + \delta^{l}_y)\\
                        (x^{l+1}_{k}, y^{l+1}_{k}) = (x^{l}_{k}, y^{l}_{k}),\quad \text{if } k \notin \{ i_l, j_l \}
                        \end{cases} \\
                        & 1 \leq i^l \leq N, 1 \leq j^l \leq N, i^l \neq j^l,\quad \forall 1 \leq l \leq L
\end{split}
\end{equation}

\subsection{Swap Compression}\label{sec:compress}
\begin{figure}[htb]
\centering
    \includegraphics[width=\columnwidth]{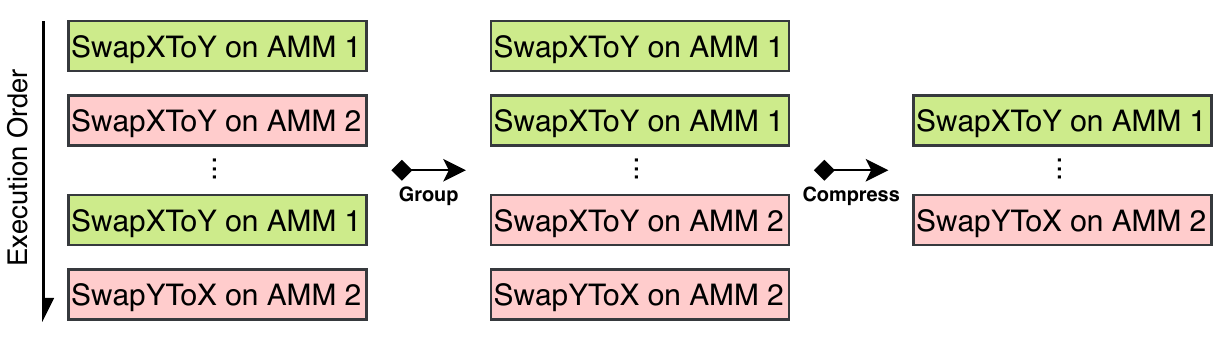}
    \caption{Overview of the swap compression process.}
    \label{fig:compress}
\end{figure}

To minimize the required transaction fees of a swap, we show in the following how \aamm compresses multiple transactions of an atomic swap (cf.\ Figure~\ref{fig:compress}).

\begin{tcolorbox}[arc=0pt,outer arc=0pt]
\begin{theorem}\label{theorem:batch_execution}
An optimal strategy ($S_{\text{optimal}}$) performing routing and arbitrage among $N$ \amms on market ($X,Y$) is equivalent to a batch execution strategy ($S_{\text{batch}}$). $S_{\text{batch}}$ consists of at most $N$ swaps (\TransactXY or \TransactYX). Both $S_{\text{optimal}}$ and $S_{\text{batch}}$ change the states  of \amms from $(x_1, y_1, \ldots, x_N, y_N)$ to $(x_1', y_1', \ldots, x_N', y_N')$.
\end{theorem}
\end{tcolorbox}

PROOF: Both arbitrage and routing only consist of \TransactXY and \TransactYX transactions (cf.\ Equation~\ref{eq:AMM_arbitrage}~and~\ref{eq:AMM_routXY}). Because the \amms are independent (cf.\ Property~\ref{prop:market_independence}), these transactions can be reordered into $N$ groups, where each group only consists of transactions for the same market. We can then batch the transactions within each group based on the path independence property (Property~\ref{prop:path_independence}). $\square$
 
\subsection{Limitations}
In this work, we consider \aamm in isolation from other exchanges on other blockchains or external centralized exchanges. However, asset prices realistically move outside of the regarded \amm, which may still create arbitrage opportunities, even if \aamm minimizes required arbitrages among the synchronized AMMs. Moreover, the cost of price synchronization grows with the number of AMMs that \aamm peers with and is therefore limited (cf.\ Section~\ref{sec:cost_multiple}). While in this work we only consider \amm with similar pricing formulas, we believe that \aamm is adoptable any \amm pricing formula, given the corresponding on-chain computation overhead.

\section{Evaluation}\label{sec:evaluation}
In our evaluation we rely on the blockchain states of Uni- and Sushiswap, two of the biggest on-chain DEXes capturing~$73.27\%$ of the market volume at the time of writing\footnote{\url{https://www.theblockcrypto.com/data/decentralized-finance/dex-non-custodial/dex-volume-monthly}, accessed March 2021}. Therefore, our \aamm implementation peers with two AMMs of the same pricing formula. In this scenario, the optimal strategy for both routing and arbitrage can be mathematically derived as we show in the appendix (cf.\ Section~\ref{app:implementation}). We use Uniswap as a pricing oracle to fetch the X/ETH prices for any arbitrary asset $X$\footnote{To avoid overestimating the revenue, we estimate the ETH value of $x$ amount of asset $X$ by simulating a Uni/Sushiswap trade, instead of relying on the spot price.}. We assume that the X/ETH price is zero when we cannot determine the price, thus ignoring the corresponding transaction. We adopt a price of~$2000$ USD/ETH as of~April~$2021$.

\subsection{Empirical Comparison of \amm and \aamm}
\begin{figure}[htb]
    \centering
    \includegraphics[width=\columnwidth]{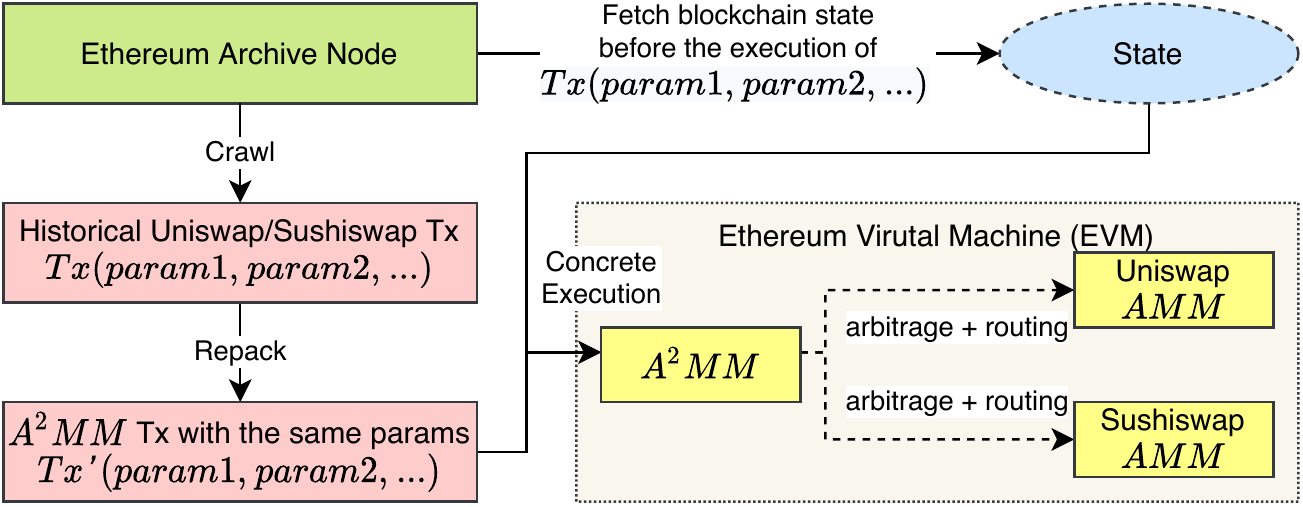}
    \caption{System model of the \aamm evaluation. We concretely execute on past blockchain data, and assume that \aamm initially does not provide a liquidity pool and rather operates through Uni- and Sushiswap.}
    \label{fig:implentation_system_model}
\end{figure}

We perform an empirical comparison between \amm and \aamm through concrete execution on past blockchain data. Our experimental setup corresponds to the system model in Figure~\ref{fig:implentation_system_model}, where we assume that we deploy an \aamm contract with a user interface while using Uni- and Sushiswap's liquidity pools. Upon receiving a swap request, \aamm derives the routing and arbitrage parameters on the fly on-chain.

We implement \aamm in 761 lines of code using Solidity v.8.2.0. The deployment of the \aamm contract costs~$2,821,822$ gas ($0.282$~ETH, $564$~USD) at a gas price of~$100$~gWei. We crawl all asset swap transactions that are sent directly to Uni- or Sushiswap from block~\ConcreteExectionStartBlock (\ConcreteExectionStartDay) to block~\ConcreteExectionEndBlock (\ConcreteExectionEndDay) (\ConcreteExectionDays days). Note that \aamm aims to reduce the number of two-point arbitrage and sandwich opportunities, where arbitrage and sandwich bots often use smart contract accounts. Because we assume that all \amm swaps initiate at the \aamm's user interface, we expect the number MEV related transactions to decrease. To avoid double-counting MEV transactions in our evaluation, we chose to only consider \amm swaps from non-smart contract accounts (i.e., EOA accounts). 

\begin{table*}[htb]
\centering
\small
\begin{tabular}{l|ll|ll|ll|ll}
\toprule
   Num. sub-swaps per swap & \multicolumn{2}{c}{Total} & \multicolumn{2}{c}{$1$}& \multicolumn{2}{c}{$2$}& \multicolumn{2}{c}{$>2$} \\
\midrule
   Transaction fee         &   \amm  & \aamm &  \amm & \aamm &   \amm & \aamm & \amm & \aamm  \\
\midrule
                   1. Swap & $127{\pm}32K$ & $136{\pm}40K$ & $121{\pm}24K$ & $128{\pm}27K$ & $182{\pm}39K$ & $208{\pm}53K$ &     $258{\pm}49K$ &       $310{\pm}74K$ \\
         2. Swap + Routing &             - & $164{\pm}60K$ &             - & $145{\pm}41K$ &             - & $225{\pm}62K$ &                 - &       $353{\pm}82K$ \\
       3. Swap + Arbitrage & $378{\pm}96K$ & $228{\pm}68K$ & $345{\pm}73K$ & $203{\pm}49K$ & $450{\pm}96K$ & $282{\pm}69K$ &    $574{\pm}103K$ &       $390{\pm}82K$ \\
\bottomrule
\end{tabular}
\caption{
Empirical transaction fee comparison between the \amm and \aamm model. We consider three cases: \emph{(i)} without routing/arbitrage \emph{(ii)} with routing, and \emph{(iii)} with arbitrage. 
}
\label{tab:fees}
\end{table*}

\point{Transaction Fees When Peered With Two \amms} \aamm requires more computation for a single swap than a standard \amm, because \aamm derives the routing and arbitrage parameters across \amms on-chain. A natural question is how much more expensive \aamm's execution ends up when compared to an \amm, when it peers with Uni-/Sushiswap. Table~\ref{tab:fees} presents our concrete execution results. For a swap without arbitrage nor routing, we find that on average, liquidity takers pay~$\GasRatioAAMMSwap$ higher fees for a swap on \aamm vs.\ \amm. For a swap with routing, we find that \aamm requires an excess~$\GasRatioAAMMRouting$ in terms of transaction fees compared to the average transaction fee of an \amm swap. Finally, for a swap with routing and arbitrage, \aamm's excess in transaction fees amounts to~$\GasRatioAAMMRoutingArbitrage$. We estimate that the arbitrage action by itself costs~$\GasRatioAAMMRoutingArbitrage - \GasRatioAAMMRouting = \GasRatioAAMMArbitrage$.



\begin{table*}[]
\centering
\footnotesize
\begin{tabular}{lllllcc}
\toprule
Type              & Num. of Txs(\%)    & Total Revenue                    & Avg. Revenue    & Avg. \aamm Fee   & Avg. $\frac{\text{Revenue}}{\text{Fee}_{\text{\aamm}}}$  & Avg. $\frac{\text{Revenue}{-}\text{Excess Fee}}{\text{Fee}_{\text{\amm}}}$ \\ \midrule
ETH$\rightarrow$Tokens          &     $860.9K (53\%)$   &  $12.0K$ETH($24M$USD) &  $0.01$ETH($28$USD) & $0.02$ETH($40$USD) &   $70\%$ & $62\%$ \\
Tokens$\rightarrow$Tokens       &     $494.2K (30\%)$   &  $17.7K$ETH($35M$USD) &  $0.04$ETH($72$USD) & $0.03$ETH($56$USD) &  $127\%$ & $137\%$ \\
Tokens$\rightarrow$ETH          &     $271.5K (17\%)$   &    $3.0K$ETH($6M$USD) &  $0.01$ETH($22$USD) & $0.02$ETH($32$USD) &   $69\%$ & $60\%$ \\
Total                           &  $1,626.6K (100\%)$   &  $32.7K$ETH($65M$USD) &  $0.02$ETH($40$USD) & $0.02$ETH($43$USD) &   $93\%$ & $90\%$ \\
\midrule
Routing                         &  $588.0K (36\%)$      &  $5.2K$ETH($10M$USD)  &  $0.01$ ETH($18$ USD) &  $0.02$ETH($36$USD) &  $48\%$    & $39\%$ \\
Arbitrage                       &  $561.7K (35\%)$      &  $27.5K$ETH($55M$USD) &  $0.05$ ETH($98$ USD) &  $0.03$ETH($60$USD) &  $164\%$   & $203\%$ \\
\bottomrule
\end{tabular}
\caption{Revenue and cost of our concrete execution on past blockchain data, replaying previous Uni/Sushiswap transactions on \aamm (i.e., the $\delta_x$ for \TransactXY remains unchanged, cf.\ Equation~\ref{eq:AMM_transact_y_for_x}). We adopt a price of~$2000$ USD/ETH as of~April~$2021$. Excess fee is the additional transaction fee \aamm costs when compared with an \amm (i.e., $\text{Fee}_{\text{\aamm}}{-}\text{Fee}_{\text{\amm}}$).}
\label{tab:revenue}
\end{table*}



\point{Extractable Arbitrage/Routing Revenue of \aamm}
In the following, we quantify the income potential from \aamm's design, as arbitrage is known to yield positive incoming from synchronizing prices. Routing provides better swap asset prices by sourcing several liquidity pools simultaneously. We proceed to measure both the positive income from arbitrage and the price advantage from routing, allowing us to offer an objective view of the costs of using \aamm compared to an \amm.

Our results suggest (cf.\ Table~\ref{tab:revenue}) that within~\ConcreteExectionDays days of blockchain data, in total~\AAMMArbRouteNum (\AAMMArbRouteRatio) of the executed \aamm transactions perform either arbitrage and/or routing, extracting a total of~\AAMMArbRouteRevenueInETH~ETH (\AAMMArbRouteRevenueInUSD~USD). Due to this positive income and the routing price advantage, in expectation, \aamm reduces transaction fees by an average of~\AverageTransactionFeeReduction compared to a standard \amm swap.

\begin{table}[]
\centering
\scriptsize
\begin{tabular}{llllllllll}
\toprule
Percentile      &     10 &    20 &    30 &    40 &    50 &    60 &   70 &   80 &    90 \\
Profit(USD) &  -20.3 &  -9.9 &  -3.8 &  -1.8 &  -1.0 &  -0.4 &  0.4 &  4.3 &  31.0 \\
\bottomrule
\end{tabular}
\caption{Percentile analysis of individual swaps' profit from our prototype implementation. We assume that traders pay the same transaction fee price (gas price) in our concrete execution. While only~$37\%$ of the swaps realize a positive profit, \aamm in expectation provides a better price than an \amm.}
\end{table}

\subsection{Two-point Arbitrage Overhead}
While \aamm only mitigates two-point arbitrage overhead, Qin~\etal~\cite{qin2021quantifying} show that historically~$41\%$ of the on-chain arbitrages are two-point arbitrages. Therefore, we estimate that \aamm will decrease about~$41\%$ of the on-chain and network overhead caused by arbitrage bots, helping to reduce the stale block rate, thus increasing blockchain consensus security.

In the following, we quantify both the on-chain and network layer overhead for the past two-point arbitrages between Uni- and Sushiswap to test the above intuition. 
\point{Block-space Overhead Heuristics} 
In the following we use $B_i$ to denote a block with height $i$, and $tx^k_i$ to denote a transaction mined within block $b_i$ at index $k$. We use $f_S(B_{i}) \mapsto S$ to denote the blockchain state after executing all transactions in block $B_{i}$. We use the function $f_S(B_{i}, tx1, \ldots, txN) \mapsto S$ to denote the blockchain state after iteratively applying transactions $tx1, \ldots, txN$ in the exact order on the blockchain state $S(B_{i})$. In other words, if there are $l$ transactions in block $B_{i}$, then $S(B_{i}) = S(B_{i-1}, tx^0_i, \ldots, tx^l_i)$. We use the function $f_A(s \in S, tx \in TX) \mapsto bool$ to classify whether a transaction $tx$ successfully performs an arbitrage at blockchain state $s$ (cf.\ Section~\ref{sec:arbitrage_heuristics} in Appendix).

We classify transaction $tx^k_i$ as an block-space overhead caused by front-/back-running arbitrages if Heuristic C1, and one of Heuristics C2a and C2b are satisfied. 

\begin{description}
\item[Heuristic C1:]
If transaction $tx^{k}_i$ performs a successful arbitrage, then $tx^{k}_i$ is not classified as an block-space overhead (cf.\ Equation~\ref{eq:oh1}).
\begin{equation}\label{eq:oh1}
    f_A(S(B_{i-1}, tx^0_i, \ldots, tx^{k-1}_i), tx^{k}_i) = \text{false}
\end{equation}
\item[Heuristic C2a (Front-running):] We test, if a re-positioning of $tx^{k}_i$ as the first transaction in each of the previous five blocks (i.e.\ a $1$-minute time window), would make $tx^{k}_i$ a successful arbitrage transaction. This test allows us to classify whether $tx^{k}_i$ is a failed front-running arbitrage overhead (cf.\ Equation~\ref{eq:oh2}).
\begin{equation}\label{eq:oh2}
    f_A(S(b_{j}), tx^{k}_i) = \text{true}, \text{where} (i-5) \leq j \leq (i-1)
\end{equation}

\item[Heuristic C2b (Back-running):] By iterating backwards through the transactions of the last~$5$ blocks, starting at $tx^{k}_i$'s position, we sequentially interleave $tx^{k}_i$ after each $tx^{l}_j$, and test through concrete execution, whether $tx^{k}_i$ yields an arbitrage profit. This test allows us to identify whether a transaction attempted an arbitrage operation (cf.\ Equation~\ref{eq:oh3}).

\begin{equation}\label{eq:oh3}
\begin{aligned}
                & f_A(S(b_{j}), tx^{l}_j, tx^{k}_i) = \text{true}  \\
\text{where:\ } & (i-5) \leq j \leq (i-1), \qquad tx^{l}_j \neq tx^{k}_i. \\
                & \text{gas price of } tx^{k}_i \leq \text{gas price of } tx^{l}_j
\end{aligned}
\end{equation}
\end{description}

\point{Network Layer Overhead Heuristics}
We classify a transaction $tx$ as a network layer overhead targeting a successful arbitrage transaction $tx_{arb}$, if the following three heuristics (N1, N2 and N3) are satisfied. Note that while a transaction may propagate on the P2P network, that transaction doesn't necessarily appear in the blockchain. The helper function $f_{\text{P2P}}(x \in {TX, B})$ returns the time of a transaction or block's first known appearance on the P2P network.

\begin{description}
\item[Heuristic N1:] N1 tests whether a transaction on the P2P layer is a failed arbitrage attempt. Given an identified successful on-chain arbitrage transaction $tx_{arb}$, we replace $tx_{arb}$ iteratively with each transaction $tx$ recorded by our P2P network node. If $tx$ yields an arbitrage profit under concrete execution, we classify $tx$ as a failed network layer arbitrage attempt, caused by either GPA or BRF.

Heuristics N2 and N3 attempt to narrow down the issuance time of an arbitrage transaction $tx$. If both the two tests in Heuristics N2 and N3 are satisfied, we then classify $tx$ as a network layer overhead transaction targeting $tx_{arb}$.

\item[Heuristic N2:] N2 tests the lower time of appearance of an arbitrage attempt. We find the earliest appearance of all $tx_{arb}$ related transactions on the P2P network, such as: \textit{(i)} the transaction that $tx_{arb}$ attempts to front-/back-run (denoted as $tx^{\text{victim}}_{tx_{arb}}$), and \textit{(ii)} all failed block-space overhead transactions competing with $tx_{arb}$ (denoted as $TX^{\text{overhead}}_{tx_{arb}}$). 
For each network layer overhead transaction $tx$, we test if $tx$ is discovered after the earliest appearance of all $tx_{arb}$ related transactions on the P2P network (cf.\ Equation~\ref{eq:no2}).

\begin{equation}\label{eq:no2}
\begin{aligned}
& t_{tx_{arb}} = f_{\text{P2P}}(tx) < min(\{f_{\text{P2P}}(x)\}), \text{where } \\
& x \in tx_{arb} {\cup} TX^{\text{overhead}}_{tx_{arb}} {\cup} tx^{\text{victim}}_{tx_{arb}} {\cup} tx^{\text{target}}_{tx_{arb}}
\end{aligned}
\end{equation}

\item[Heuristic N3:] N3 tests the upper time of appearance of an arbitrage attempt. For each network layer overhead transaction $tx$, we test if $tx$ is discovered before $tx_{arb}$ is mined (cf.\ Equation~\ref{eq:no3}).

\begin{equation}\label{eq:no3}
\begin{aligned}
f_{\text{P2P}}(tx) > f_{\text{P2P}}(b), \text{where } tx_{arb} \text{ is mined in } b
\end{aligned}
\end{equation}
\end{description}

\begin{figure}[htb]
\centering
    \includegraphics[width=\columnwidth]{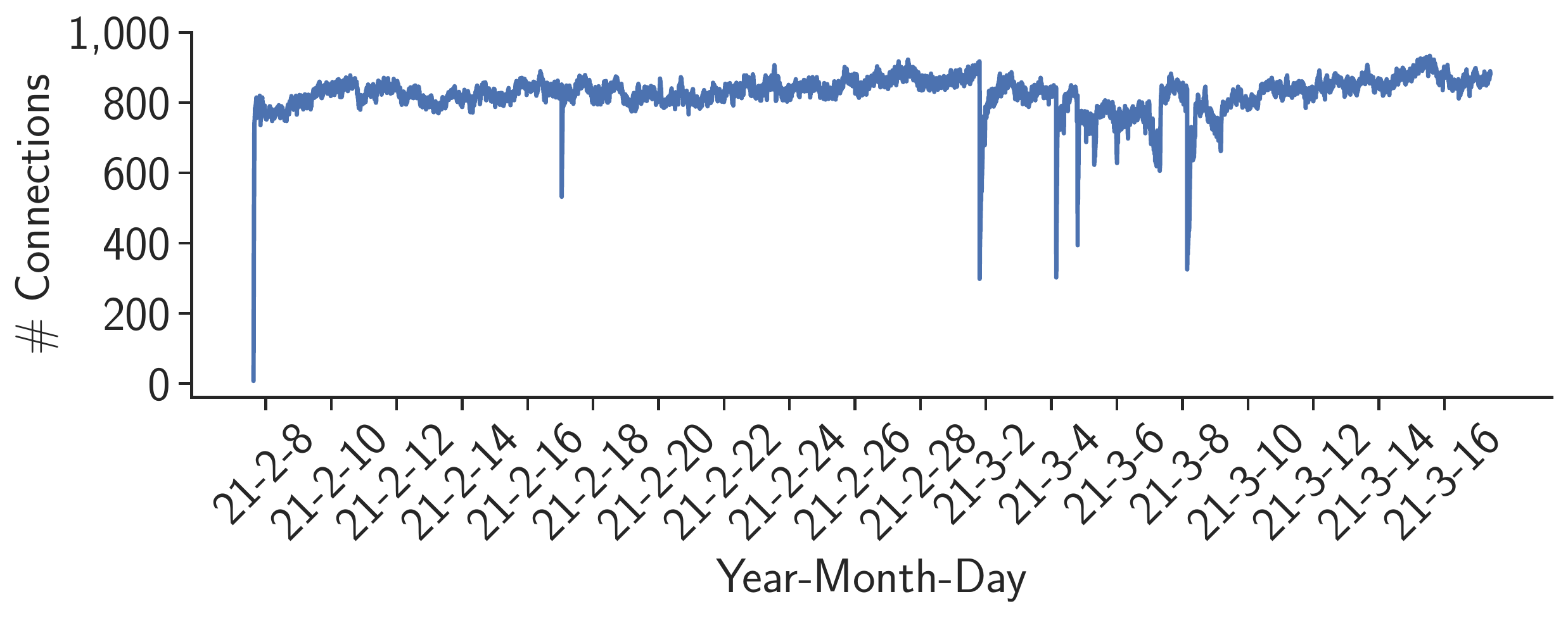}
    \caption{Number of connections of our modified geth node while listening on the Ethereum P2P network from block~$11,813,201$ (Feb-$08$-$2021$) to block~$12,055,081$ (Mar-$17$-$2021$).}
    \label{fig:connections}
\end{figure}

\point{Empirical Results} To quantify the amount of P2P network layer overhead caused by two-point arbitrages, we modify the Ethereum geth client to store all transactions received on the P2P network layer over~$213,538$ blocks ($36$ days) from block~$11,813,201$ (Feb-$08$-$2021$) to block~$12,055,081$ (Mar-$17$-$2021$). Intuitively, the number of transactions the Ethereum node can observe increases with the number of peer connections, the network bandwidth, and the computation power of the machine. Our geth client operates on a Ubuntu~$20.04.1$ LTS machine with AMD Ryzen Threadripper~$3990X$ ($64$-core,~$2.9$ GHz),~$256$~GB of RAM and~$4\times2$ TB NVMe SSD in Raid~$0$ conﬁguration. We limit the geth client to at most~$1,000$ connections with other Ethereum peers instead of the default of~$50$ peers (cf.\ Figure~\ref{fig:connections}). We captured in total~$246$B transaction propagation messages from~$81,736$ unique peers originating from $63,744$ unique IP addresses and $2,859,833$ unique IP:Port combinations.

\begin{table}[]
\small
\begin{tabular}{llll}
\toprule
Position & Front-running &   Back-running &          Total \\
\midrule
Same block      &  13,460(73\%) &   117,728(90\%) &   131,188(88\%) \\
After 1 block  &   2,876(16\%) &     7,657(6\%)  &    10,533(7\%)  \\
After 2 blocks &     909(5\%)  &     2,118(2\%)  &     3,027(2\%)  \\
After 3 blocks &     465(3\%)  &     1,353(1\%)  &     1,818(1\%)  \\
After 4 blocks &     419(2\%)  &       860(1\%)  &     1,279(1\%)  \\
After 5 blocks &     268(1\%)  &       775(1\%)  &     1,043(1\%)  \\
\bottomrule
\end{tabular}
\caption{Statistics of the block-space overhead we detect. $95\%$ of the on-chain failed arbitrages are mined within $1$ block after the MEV arbitrage opportunity is extracted.}
\label{tab:frontrunning_overhead}
\end{table}

\begin{figure*}[htb]
\centering
    \includegraphics[width=\textwidth]{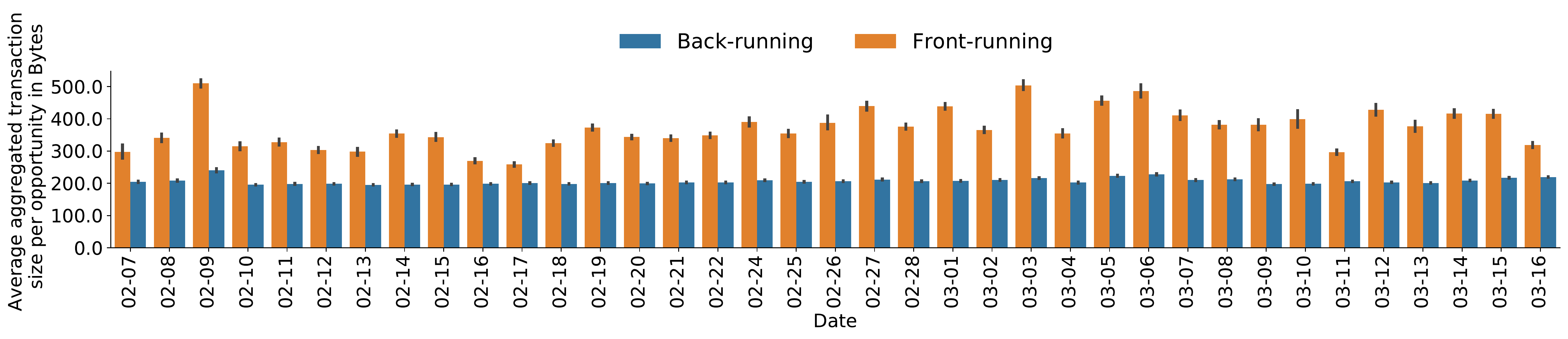}
    \caption{For every two-point arbitrage opportunity on-chain, we correlate the accumulative overhead transactions propagated on the P2P network layer. We capture front- as well as back-running transactions, covering PGA and BRF.}
    \label{fig:daily_overhead}
\end{figure*}

\begin{figure*}[htb]
\centering
    \includegraphics[width=\textwidth]{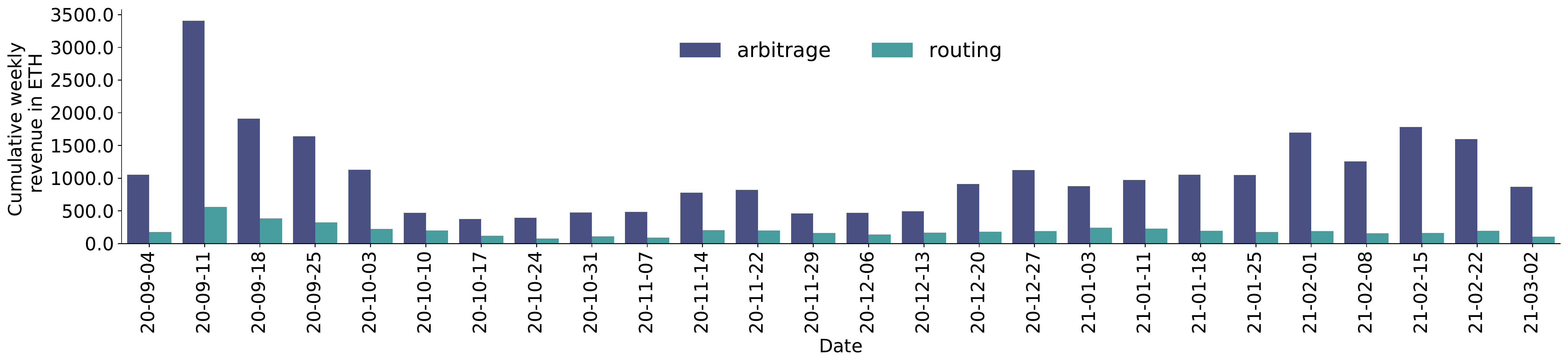}
    \caption{Cumulative weekly revenue of the \aamm implementation extracted through arbitrage and routing based on concrete execution of past blockchain data.}
    \label{fig:revenue_over_time}
\end{figure*}

\begin{description}
\item[Potential Freed up Block-Space by \aamm] Given the heuristics C1 and C2, we identify in total~$148,888$ on-chain two-point arbitrage overhead transactions (cf.\ Table~\ref{tab:frontrunning_overhead}). Surprisingly, the majority are back-running arbitrage failures ($130,491/87.64\%$). On average,~$2.5$ and~$6.0$ overhead transactions are mined on-chain for each front- and back-running opportunities, with an average gas cost of ~$193 \pm 70$K. As front-running arbitrageurs participate in PGA, front-runners pay a~$1.56 \times$ premium average gas price compared to back-runners. We use Equation~\ref{eq:gas_reduce} to quantify \aamm's on-chain reduction over $213,538$ blocks ($36$ days), where $C$ denotes the average on-chain space cost. We find that in expectation, \aamm reduces the consumed block-space by~\BlockSpaceReduction.

\begin{equation}\label{eq:gas_reduce}
C_\text{reduced} = 1 - \frac{C_{\text{A}^2\text{MM swap}}}{C_{\text{AMM swap}} + C_{\text{arbitrage}} + C_{\text{block-space overhead}}}
\end{equation}

\item[Potential Network Overhead Reduced by \aamm] Given the heuristics N1, N2 and N3, we identify~$105,960$ network overhead transactions, where the majority ($400,471/89.4\%$) are caused by back-running arbitrageurs. On average,~$10.5$ and~$27.4$ network overhead transactions are issued for every front-/back-running arbitrage opportunity, which corresponds to a factor of~$4.2\times, 4.6\times$ more than the block-space overhead. When an arbitrage opportunity appears, we measure that the off-chain overhead transactions sum to an average of~\AverageBytesNetworkOverhead kb per block, which is around~\AverageBytesNetworkOverheadPercentage of a block's size on the~1st of March~2021\footnote{\url{https://etherscan.io/chart/blocksize}}.
\end{description}

\point{Limitations} 
Our evaluation may consist of false negatives. For example, an overhead arbitrage transaction may be dropped in the asynchronous P2P network before reaching our network node. The blockchain overhead statistics we report should therefore only be regarded as a lower bound of the actual network overhead. Note that a transaction is only classified as an overhead, if it does perform a two-point arbitrage during concrete execution. Therefore, our overhead evaluation suffers from no false positives.

\section{Security Implications of \aamm}\label{sec:security-implications}

In the following, we quantitatively outline the relevant security improvements \aamm provides on the blockchain consensus.

\subsection{Stale Block Rate Simulation}
This section simulates the P2P network of four blockchains (Ethereum, Bitcoin, Litecoin, and Dogecoin) to estimate quantitatively the relationship between the stale block rate and the miner bandwidth. To capture the block propagation in the P2P network, we extend our system model from Section~\ref{sec:aamm}. The asynchronous nature of blockchain P2P propagation is extensively studied by related works~\cite{ersoy2018transaction, decker2013information, gervais2016security, gencer2018decentralization, zhou2020high}, on which we build upon. 

Various factors influence block propagation, including the number of miners, the network topology, the peer internet latency, bandwidth, and overall network congestion. To ease our experiments and operate under the best network connectivity, we assume that the miners create direct point-to-point relations among themselves. Consequently, the number of sporadic network nodes, the network topology, intermediate devices (relay nodes, switches, and routers), and the TCP congestion management are all abstracted. We approximate the block propagation duration by dividing the block size over the bandwidth and adding the communication latency. 
To parameterize a realistic block size distribution in our simulations, we assume that the block size follows a normal distribution, where the mean and variance are derived using~$90$ days of blockchain data (cf.\ Table~\ref{tab:blockchain_stats}) \footnote{\url{https://bitinfocharts.com}, accessed Apr 2021}. To capture latency distribution, we apply the mean percentile statistics~\cite{kim2018measuring, gencer2018decentralization, zhou2020high} and use linear interpolation to estimate the underlying cumulative probability distribution (cf.\ Table~\ref{tab:network_statistics} in the Appendix). We only consider the hashing power of the top $10$ miners for each blockchain (cf.\ Table~\ref{tab:hashing_power} in the Appendix), and assume that miners have a symmetrical upload and download bandwidth.

\begin{table}[]
\centering
\footnotesize
\begin{tabular}{lllll}
\toprule
                         & Ethereum  & Bitcoin    & Litecoin     & Dogecoin    \\
\midrule
Block interval (min)     & 0.223     & 9.474      & 2.59      & 1.07   \\
Block size mean (kB)     & 44.0      & 863.8      & 61.1      & 15.9    \\
Block size std (kB)      & 3.0       & 25.0       & 33.4      & 14.9   \\
\bottomrule
\end{tabular}
\caption{Blockchain parameters we use to simulate the P2P network for Ethereum, Bitcoin, Litecoin and Dogecoin. These statistics are measured using~$90$ days of blockchain data, from~Jan-$09$-$2021$ to~Apr-$09$-$2021$.
}
\label{tab:blockchain_stats}
\end{table}

\begin{figure}[htb]
\centering
    \includegraphics[width=\columnwidth]{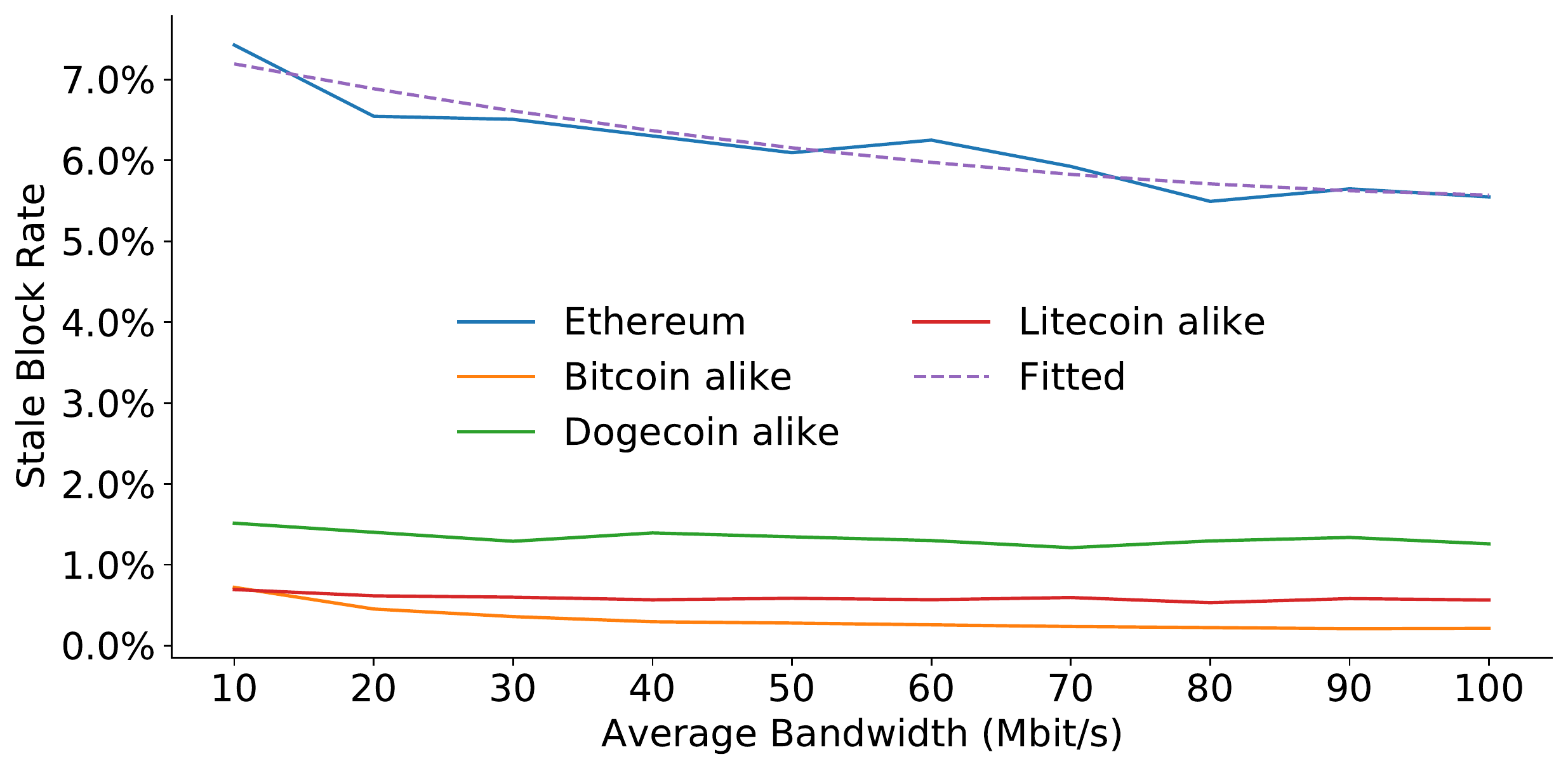}
    \caption{Simulated stale block rate given the average P2P network bandwidth for Ethereum, Bitcoin, Litecoin and Dogecoin. We fit a least square regression line for Ethereum ($0.000158 \times \text{bandwidth}^2 - 0.03541 \times \text{bandwidth} + 7.531$).}
    \label{fig:stale}
\end{figure}

MEV network overhead deteriorates the miners' P2P bandwidth and hence increases the blockchain's stale block rate. The most significant arbitrage back-running in terms of overhead we capture amounts to a total of~$1.92$ Mb data from repeated transactions within one block interval ($13$ seconds). Note that the total amount of data miners receive is amplified as the number of connected peers increases. For instance, if miners have an initial upload and download speed of $70$Mbit/s, and overhead transactions are propagated back and forth~$200$ times, the average bandwidth will decrease to~$40.5$Mbit/s \footnote{$\frac{1.92 \text{Mb}}{13 \text{s}} \times 200 = 40.5\text{Mbit/s}$}. Based on our $2$nd degree least-square polynomial fitting, this decrease in bandwidth leads to a~$0.53\%$ increase in stale block rate (cf.\ Figure~\ref{fig:stale}).

Note that miners have an incentive to connect with as many nodes as possible to minimize the risks of eclipse attacks~\cite{heilman2015eclipse}, while the network layer overhead is amplified with the number of connections. We leave the estimation of such network overhead amplification factor to future works.

\subsection{Sandwich Attack Mitigation}
Sandwich attacks are not profitable if the victim's input amount remains below the MVI~\cite{zhou2020high}. This threshold depends on the \amm pricing formula, the total underlying pool liquidity, as well as the trader's slippage configuration. The MVI threshold for instance increases if the market liquidity increases.

By routing the trading volume onto multiple \amm exchanges, \aamm aggregates the MVI thresholds among the underlying liquidity pools. In the simple case, where two \amm markets have the same liquidity and pricing formula, \aamm's accumulative MVI threshold is~$2\times$ the MVI of a single \amm.

\subsection{Back-run Flooding Overhead Reduction}
We observe \emph{back-run flooding} on the P2P network, where MEV bots broadcast multiple similar back-running transactions for a single MEV opportunity (cf.\ Table~\ref{tab:overhead_tax}). It appears that BRF may increase the success rate of back-running. For instance, each of the flooding transactions is likely to follow a different network propagation path in the asynchronous P2P network, which could increase the likelihood of a swift miner reception. While we find that $88.80\%$ of the successful arbitrage transactions are accompanied by BRF, we cannot provide quantitative insights to what degree BRF improves the success-rate of back-running.

\begin{table}[]
\small
\begin{tabular}{lll}
\toprule
                                & \textbf{Front-running}        & \textbf{Back-running}     \\
\midrule
\textbf{Strategy}               & Priority Gas Auctions         & Back-run Flooding         \\
\textbf{Target State}           & Confirmed Block State         & Pending Block State       \\
\midrule
\multicolumn{3}{c}{\textbf{Overhead}} \\
\midrule
\textbf{On-chain?}              & Yes                           & Yes                       \\
\textbf{Network?}               & Yes                           & Yes                       \\\bottomrule
\end{tabular}
\caption{If an MEV bot acts on a confirmed block state, it performs PGA against other competing front-running bots. If an MEV bot acts on a pending block state, we observe back-run flooding. Both GPA and BRF cause on-chain and network layer overheads.}
\label{tab:overhead_tax}
\end{table}

To quantify the network layer overhead, we identify past arbitrages on-chain and correlate the dropped transactions on the P2P network provided by our network listening node. We find that one of the most amplified flooding events entails~$358$ transactions on the network layer for a single arbitrage opportunity. These back-running transactions are identical, except the last byte of the transaction message, floods~$65.7$kb of data traversing the P2P network. This is equivalent to~$1.5\times$ the average block size on the~1st of March~2021\footnote{\url{https://etherscan.io/chart/blocksize}}. Only one of these transactions is confirmed on-chain\footnote{\etherscanTx{0x49bc22c9c45d31064f3cf7f7bd5e1494439603d4f6e809b0a715bc08d1b585c8}}, classified as a failed arbitrage attempt by us. The remaining~$357$ transactions have a conflicting nonce with the confirmed transaction, and therefore discarded. We observe that back-run flooding is comparatively cheap because bots issue conflicting MEV transactions (e.g., with the same nonce), while only one transaction is mined.

\section{Arbitrage/Routing Among $N$ \amms}\label{sec:cost_multiple}
In this section, we shed light on the performance of \aamm when peered with N \amm markets (abbreviated as~$N$-\aamm). While Section~\ref{app:implementation} in the appendix provides an optimal arbitrage strategy of~$2$-\aamm, Algorithm~\ref{alg:suboptimal_arbitrage} presents our sub-optimal two-point arbitrage strategy for~$N$-\aamm, where~$N>2$. Intuitively, our strategy starts with the two \amms offering the best and worst prices, and gradually narrows the price gap through arbitrage. Along this narrowing process, if the prices of a group of \amms are synchronized, we aggregate their liquidity and treat them virtually as a single exchange. Executing a swap on a virtually aggregated exchange is equivalent to performing routing, where the trade volume is routed to each of the underlying \amm based on their liquidity. Our strategy, therefore, translate the arbitrage problem of~$N$-\aamm into~$2$-\aamm sub-problems. 

\begin{figure}[htb!]
    \centering
    \includegraphics[width=\columnwidth]{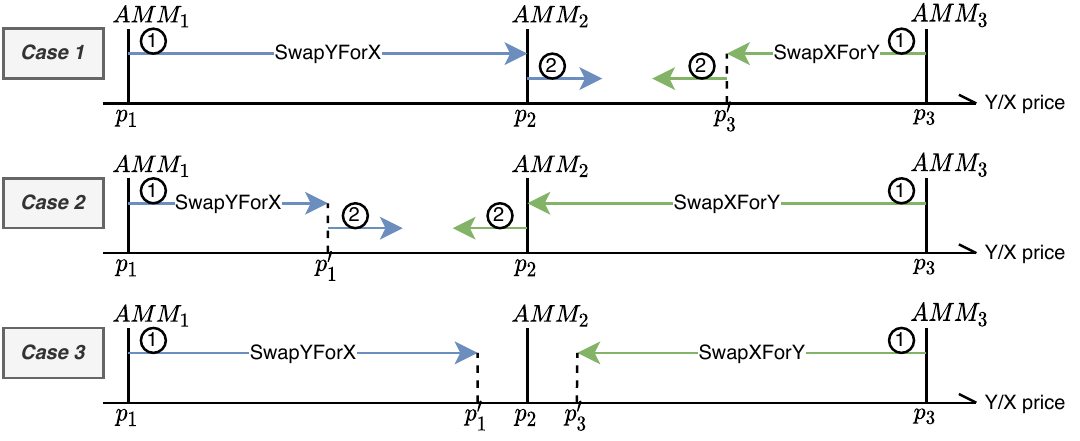}
    \caption{Visualization of the two-point arbitrage process among three \amms, which performs at most~$3$ swaps (either \TransactYX or \TransactXY). In the first case, the liquidity from $\amm_1$ and $\amm_2$ are aggregated for \TransactYX. The second case is symmetric to the first. In the third case, the arbitrage does not trigger \amm aggregation.}
    \label{fig:3AMMArbitrage}
\end{figure}

To ease the reader's understanding, we visualize the arbitrage process among three \amms in Figure~\ref{fig:3AMMArbitrage}. The three \amms we consider have prices sorted in ascending order ($p_1$, $p_2$, and $p_3$ respectively). Our algorithm hence performs arbitrage by considering only \amm~$1$ and~$3$ first. As the price gap narrows, we can encounter three different cases. In the first case, the price of \amm~$1$ increases from $p_1$ to $p_2$, which is synchronized with the price of \amm 2. Our algorithm then aggregates these two exchanges, and continues the arbitrage process between the newly aggregated virtual \amm and \amm~$3$. The second case is the symmetric to the first case, where the price of \amm~$3$ falls from $p_3$ to $p_2$, and our algorithm aggregates \amm~$2$ and~$3$. In the last case, (due to fees) the prices of all three \amms are not synchronized, therefore we do not aggregate any \amms.

Table~\ref{tab:3ammfees} shows the cost of Algorithm~\ref{alg:suboptimal_arbitrage} among $N$ constant product \amms. We estimate that the transaction cost of $3$-\aamm is~$1.7\times$ the cost for $2$-\aamm. We estimate the costs by applying linear interpolation based on our empirical cost evaluation from Table~\ref{tab:fees}.


\begin{table}[h!]
\footnotesize
\centering
\begin{tabular}{l|lll}
\toprule
                                                & \multicolumn{3}{c}{Number of synchronized \amms}                                                                                 \\
Type                                            & $2$                           & $3$                       & $N>3$                                     \\
\midrule
\multicolumn{4}{l}{Number of} \\
Arbitrage computation                                                 & $1$                       & $2$                       & $N{-}1$                                   \\
Synchronize volume                                                & $0$                       & $1$                       & $2N{-}5$                                  \\
Swaps                                                                & $2$                       & $3$                       & $N$                                       \\
\midrule
\multicolumn{4}{l}{Cost over \amm average swap cost} \\
Routing to one \amm                                                  & \underline{$\GasRatioAAMMRouting$}    & \underline{$\GasRatioAAMMRouting$}    & \underline{$\GasRatioAAMMRouting$}                      \\
Arbitrage computation                                                & \underline{$\GasRatioAAMMArbitrage$}  & $84.84\%$                 & $\GasRatioAAMMArbitrage {\times} (N{-}1)$            \\
Threshold computation                                                & N/A                       & \underline{$\GasRatioAAMMRouting$}    & $\GasRatioAAMMRouting {\times} (2N{-}5)$           \\
Swap execution                                                       & $100\%$                   & $150\%$                   & $200\%$                                   \\
Total cost                                                           & $160.22\%$                & $270.44\%$                & $\!\begin{aligned}[t]
                                                                                                                                            &217.80\% \\
                                                                                                                                            &{+}\GasRatioAAMMArbitrage{\times}(N{-}1) \\
                                                                                                                                            &{+}\GasRatioAAMMRouting{\times}(2N{-}5) \\
                                                                                                                                        \end{aligned}$\\
\bottomrule
\end{tabular}
\caption{Cost prediction of performing two-point arbitrages among multiple \amms when a user sells asset X to purchase asset Y. The underlined cost ratios are taken from our two-point arbitrage evaluation (cf. Table~\ref{tab:fees}). The synchronize volume quantifies the amount of trading volume required to synchronize the asset price among multiple \amms. We estimate that the cost of synchronize volume is similar to the cost of a single optimal routing.}
\label{tab:3ammfees}
\end{table}

\section{Related Work}\label{sec:related-work}

\point{AMM}
The literature proposes various blockchain-based exchange models covering limit order book models~\cite{kyber2019,oasis2019,idex2019}, auctions~\cite{dutchx2019}, payment channel~\cite{luo2019payment} and trusted hardware~\cite{bentov2019tesseract} were proposed in the literature. Uniswap is to date the most actively used constant product AMM, while alternative weighted AMMs emerged~\cite{balancerexchange}.

\point{Arbitrage}
Identifying arbitrage opportunities is extensively studied in traditional, centralized finance (or CeFi)~\cite{cai2003approximation,cui2020arbitrage}. One common methodology is to create a graph of all pairwise assets that can be traded to use a greedy search strategy, such as the Bellman-Ford-Moore algorithm, to search the trading space. For instance, the Bellman-Ford-Moore algorithm operates with a complexity of $O(E*V)$ in a graph of $E$ edges and $V$ vertices. Such a greedy search methodology aims to create a circular, profitable trading opportunity. Greedy search approaches are restricted to actions such as trade asset $X$ for $Y$. However, because a greedy search algorithm only follows the locally optimal choice at each action, it might fail to explore and find profitable trading strategies. Zhou et al.\ show two mechanisms to automatically discover profitable arbitrage opportunities in the intertwined DeFi contract graph~\cite{zhou2021just}. Bartoletti et al.\ distill fundamental structural and economic aspects of AMMs, and in particular discuss the arbitrage problem~\cite{bartoletti2021theory}.

\point{Front-Running and Miner Extractable Value} Bonneau~\etal~\cite{bonneau2016buy}, introduce the concept of bribery attack, which incentivizes miners to adopt a blockchain fork instead of the longest chain. Daian~\emph{et al.}~\cite{daian2019flash}, introduce the concept of gas price auctions (PGA) among trading bots as well as the concept of MEV. MEV widens the variance of block rewards, encourages both bribery and under-cutting attacks~\cite{bonneau2016buy,carlsten2016instability}. The literature captures front-running by allowing a ``rushing adversary'' to interact with a protocol~\cite{beaver1992cryptographic}. Previous studies~\cite{baron2019risk,zhou2020high} suggest that HFT performance is strongly associated with latency and execution speed. The (financial) high-frequency trading (HFT) literature~\cite{angel2013fairness,menkveld2016economics} explores several trading strategies and their economic impact, such as arbitrage, news reaction strategies, etc. in traditional markets. Most of the traditional finance market strategies are applicable to AMM and decentralized exchanges~\cite{daian2019flash,angeris2019analysis,qin2020attacking,qin2021quantifying,zhou2020high,zhou2021just}.

\point{Eclipse Attacks}
Strategically placed blockchain network nodes may control when and if miners receive transactions, affecting the transaction execution time~\cite{marcus2018low,henningsen2019eclipsing,gervais2015tampering,heilman2015eclipse}.

\point{Malpractices on Exchanges}
Malpractices on financial exchanges is a well-studied domain. Besides the traditional market manipulation techniques~\cite{jarrow1992market} (such as cornering, front-running, and pump-and-dump schemes), previous works~\cite{lin2016new} studies techniques such as spoofing, pinging, and mass misinformation, which leverage, e.g., social media, artificial intelligence, and natural language processing. Techniques were shown to deceive HFT algorithms~\cite{arnoldi2016computer}. To counterbalance this inherent trust, regulators conduct periodic and costly manual audits of banks, brokers, and exchanges to unveil potential misbehavior. Because DEXes operate under weak identities and censorship resilience (from both the creators, users, and miners), regulators may face challenges to impose anti-money laundering legislation.

\section{Discussion}\label{sec:discussion}
We hope that our work engenders a wider corpus of orthogonal blockchain application designs which take into account the nature of the underlying ledger. We would like to emphasize that our work is based on a non-optimized prototype implementation which can likely be improved through additional engineering efforts. 

Our evaluation shows that \aamm does lower the required exchange block-space by~\BlockSpaceReduction. As such, \aamm classifies as a scaling solution for both the network as well as the blockchain layer. While most existing backward-compatible scaling solutions such as payment channels, off-chain hubs, etc~\cite{gudgeon2019sok} provide weaker security guarantees, \aamm inherits as a decentralized application the native blockchain security properties, and moreover improves the security of the blockchain consensus as shown in this paper.

\section{Conclusion}\label{sec:conclusion}
By means of the realization that one blockchain should only operate at most one \amm exchange, we design a novel \aamm exchange, which allows exchange users to atomically extract Miner Extractable Value, instead of leaving such opportunity to others. We show how \aamm can avoid two-point arbitrage MEV overhead on the P2P network and the blockchain transaction space. Reducing such overhead allows to strengthen the blockchain's consensus security, without resorting to centralized relayer which undermine the very reason permissionless blockchains exist.

While \aamm inherently takes advantage of the atomic nature of blockchain transactions for arbitrage and routing, our proposal can serve as inspiration to design further MEV-friendly DeFi protocols, e.g., for liquidations in lending markets. We hope that our work provides insights into a previously unconsidered and orthogonal design space for secure DeFi protocols which sustainably recognize the decentralized characteristics of permissionless ledgers.

\bibliographystyle{plain}
\bibliography{references.bib}

\begin{appendices}

\section{MEV Overhead Taxonomy}
In the following, we provide a high-level taxonomy of the different sources of technical overhead introduced by MEV opportunities. We primarily differentiate between MEV opportunities on \emph{(i)} the not-yet-confirmed network layer state and on \emph{(ii)} the confirmed blockchain state. Note that we ignore the existence of blockchain forks for simplicity. We consider the MEV aware bots to not being miners, while the following reasoning also applies to miners extracting MEV.

\point{Confirmed Block State MEV}
An MEV extraction bot can choose to only act on a confirmed blockchain state, i.e., once a block is mined. Once a block at height $i$ is received by the bot, the bot attempts to front-run all other transactions in the next block $i+1$. Confirmed state front-running is destructive, meaning that the bot bears no consideration to the subsequent transactions in the same block~\cite{qin2021quantifying,eskandari2019sok}.

\point{Unconfirmed Block State MEV}
An MEV extraction bot may observe the unconfirmed blockchain transactions and anticipate how the next miner would order these transactions within a block (e.g., given the paid transaction fees). Based on the anticipated transaction ordering, the arbitrageur then verifies whether new MEV opportunities surfaced. If an MEV opportunity is found, the bot ideally issues a back-running arbitrage transaction as an exploit
~\cite{qin2021quantifying,eskandari2019sok}.

\point{Priority Gas Auction (PGAs) Overhead}
PGA is the process by which MEV aware bots are competitively bidding on transaction fees to obtain a specific transaction position (usually the first) in the next block. Because a transaction in the miner's mempool can be replaced before it is confirmed, PGA bots usually emit a new transaction with higher bids to replace their previous transaction~\cite{daian2019flash,zhou2020high}. Although the replaced transaction is dropped by the network eventually after the new transaction is confirmed, the replaced transaction is still broadcasted on the network layer. Therefore PGA causes an overhead on the blockchain network layer.

\point{Block-space Overhead} 
Trading bots increasingly extract MEV with optimal parameters~\cite{qin2020attacking}, bequeathing no revenue for following MEV bot transactions, which should then either revert with an error or fail silently. We classify failed successful MEV transactions as on-chain MEV overhead.

\section{Implementation}\label{app:implementation}
In this section, we present a concrete \aamm implementation with two constant product AMM DEXes, namely Uniswap V2 and Sushiswap. Both these two exchanges follow a constant product formula, with a commission fee of $0.3\%$ (cf.\ Equation~\ref{eq:constant_product}).

\begin{equation}\label{eq:constant_product}
\small
\begin{split}
    p_{X \rightarrow Y}(x, y, \delta_x) = y - \frac{x \cdot y}{x + \delta_x \cdot (1 - 0.3\%)} \\
    p_{Y \rightarrow X}(x, y, \delta_y) = x - \frac{x \cdot y}{y + \delta_y \cdot (1 - 0.3\%)}
\end{split}
\end{equation}

In the following, we denote the Uniswap and Sushiswap ($X,Y$) market as DEX 1 and 2, where the price of \amm 1 is greater than or equal to the price of \amm 2 for \TransactXY (i.e., $\frac{y_1}{x_1} \geq \frac{y_2}{x_2}$). We use $(x_1, y_1)$ and $(x_2, y_2)$ to denote the states of DEX 1 and DEX 2 respectively.

\subsection{\RouteXY}
As we have shown in Section~\ref{sec:optimal-on-chain-swap-routing}, the optimal routing strategy is to greedily route the trading volume to \amm 1 until the prices of both markets are synchronized. After the price synchronization, the remaining volume is routed to both \amm 1 and 2, while keeping the prices the same (cf.\ Theorem~\ref{theorem:route-lowest-price}).

In Equation~\ref{eq:routing_threshold}, we compute the threshold ($\bar{\delta_x}$), such that the prices between \amms 1 and 2 will be synchronized after swapping exactly $\bar{\delta_x}$ of asset $X$ for asset $Y$ on \amm 1.

\begin{equation}\label{eq:routing_threshold}
\small
\begin{split}
    & \frac{y'_1}{x'_1} = \frac{y_2}{x_2} \Longleftrightarrow \frac{\frac{x_1 \cdot y_1}{x_1 + \bar{\delta_x} \cdot 0.997}}{x'_1 + \bar{\delta_x}} = \frac{y_2}{x_2}  \\
    \Longleftrightarrow &\bar{\delta_x} \approx \frac{1.002 (\sqrt{x_{1} y_{2} \left(2.257 \cdot 10^{-6} x_{1} y_{2} + x_{2} y_{1}\right)}-x_{1} y_{2})}{y_{2}}
\end{split}
\end{equation}

We now consider the optimal routing strategy if the prices between \amms 1 and 2 are synchronized (i.e., $\frac{y_1}{x_1} = \frac{y_2}{x_2}$). We use $q = \frac{x_1}{x_2} = \frac{y_1}{y_2}$ to denote the ratio of funds between the two DEXes. In Equation~\ref{eq:routing_param}, we compute the optimal routing ratio $k$ given that the liquidity taker trades $\delta_x$ amount of asset $X$, where we route $k \cdot \delta_x, (1-k) \cdot \delta_x$ to \amms 1 and 2 respectively.

\begin{equation}\label{eq:routing_param}
\small
\begin{split}
\frac{y'_1}{x'_1} = \frac{y'_2}{x'_2} \Longleftrightarrow & \frac{\frac{x_1 \cdot y_1}{x_1 + k \cdot \delta_x \cdot 0.997}}{x_1 + k \cdot \delta_x} = \frac{\frac{x_2 \cdot y_2}{x_2 + (1-k) \cdot \delta_x \cdot 0.997}}{x_2 + (1-k) \cdot \delta_x} \\
\Longleftrightarrow & \frac{\frac{q^2 \cdot x_2 \cdot y_2}{q \cdot x_2 + k \cdot \delta_x \cdot 0.997}}{q \cdot x_2 + k \cdot \delta_x} = \frac{\frac{x_2 \cdot y_2}{x_2 + (1-k) \cdot \delta_x \cdot 0.997}}{x_2 + (1-k) \cdot \delta_x} \\
\Longleftrightarrow & \frac{x_2 + (1-k) \cdot \delta_x \cdot 0.997}{ x_2 + \frac{k}{q} \cdot \delta_x \cdot 0.997} = \frac{x_2 + \frac{k}{q} \cdot \delta_x}{x_2 + (1-k) \cdot \delta_x} \\
\Longleftarrow      & 1-k = \frac{k}{q} \Longleftrightarrow  k = \frac{q}{1+q}
\end{split}
\end{equation}

Therefore, the optimal routing strategy routes $\bar{\delta_x}$ to \amm 1 first, such that the prices between the two exchanges are synchronized. The routing strategy then routes $\frac{q}{1+q}$ of the remaining liquidity to \amm 1, and $\frac{1}{1+q}$ to \amm 2 (cf.\ Equation~\ref{eq:routing_param}).


\subsection{\ArbitrageForX}\label{sec:evaluation_arbitrageforx}
In the following, we derive profitable arbitrage constraints among two constant product \amm exchanges (e.g., Uniswap and Sushiswap). The constraints are mathematically simple, such that a smart contract derives it at low costs on-chain. We also derive the formulas to calculate optimal two-point arbitrage parameters to maximize arbitrage revenue. Equation~\ref{eq:translated_obj} shows the specific arbitrage objective function for two constant product \amms, derived by substituting Equation~\ref{eq:constant_product} into Equation~\ref{eq:arb_obj}.

\begin{equation}\label{eq:translated_obj}
\small
    \text{maximize}\quad \delta_x' - \delta_x = p^{Y \rightarrow X}(x_2, y_2, p^{X \rightarrow Y}(x_1, y_1, \delta_x)) - \delta_x
\end{equation}

To find the optimal arbitrage parameter ($\delta_x^{optimal}$), we solve the derivative of the objective function $\frac{d}{dx}(\delta_x' - \delta_x) = 0$. Equation~\ref{eq:optimal_in} shows the only positive solution for $\delta_x^{optimal}$.

\begin{equation}\label{eq:optimal_in}
\small
\delta_x^{optimal} \approx \frac{1.003 \left(- 1000 x_{1} y_{2} + 997 \sqrt{x_{1} x_{2} y_{1} y_{2}}\right)}{997 y_{1} + 1000 y_{2}}
\end{equation}

In Equation~\ref{eq:revenue}, we substitute $\delta_x^{optimal}$ into the objective function to derive the optimal revenue.

\begin{equation}\label{eq:revenue}
\small
\begin{split}
\delta_x' - \delta_x \approx & - \frac{x_{2} y_{2}}{- \frac{0.997 x_{1} y_{1}}{x_{1} - \frac{0.997 \left(c_1 x_{1} y_{2} - c_2 \sqrt{x_{1} x_{2} y_{1} y_{2}}\right)}{997 y_{1} + 1000 y_{2}}} + 0.997 y_{1} + y_{2}} \\
& + x_{2} + \frac{c_1 x_{1} y_{2} - c_2 \sqrt{x_{1} x_{2} y_{1} y_{2}}}{997 y_{1} + 1000 y_{2}} \\
\text{where}\quad c_1 &\approx 1003.009 \\
c_2 &\approx 1000.000
\end{split}
\end{equation}

In Equation~\ref{eq:profitable}, we find the constraint for the arbitrage opportunity to be profitable, without considering transaction fees. 

\begin{equation}\label{eq:profitable}
\small
\delta_x' - \delta_x > 0 \implies \frac{y_2}{x_2} < 0.994 \frac{y_1}{x_1}
\end{equation}

We, therefore, apply Equation~\ref{eq:profitable} to verify whether arbitrage is profitable given a blockchain state and use the optimal parameters (cf.\ Equation~\ref{eq:optimal_in}) to extract the maximum revenue.

\section{Miner Hashing Power}
Table~\ref{tab:hashing_power} shows the hashing power we extract from various sources to simulate the P2P network for Ethereum\footnote{\url{https://etherscan.io/stat/miner?blocktype=blocks}, accessed Apr 2021}, Bitcoin\footnote{\url{https://btc.com/stats/pool}, accessed Apr 2021}, Litecoin\footnote{\url{https://www.litecoinpool.org/pools}, accessed Apr 2021} and Dogecoin\footnote{\url{https://explorer.viawallet.com/doge/pool}, accessed Apr 2021}.

\begin{table}[h!]
\centering
\begin{tabular}{ccccc}
\toprule
Rank  &      Ethereum  & Bitcoin & Litecoin & Dogecoin \\
\midrule
    1 &        24.3\%  &17.9\% & 16.0\% & 14.9\%\\
    2 &        19.3\%  &15.5\% & 14.4\% & 13.61\% \\
    3 &        10.4\%  &11.9\% & 14.0\% & 13.38\% \\
    4 &        5.8\%  &11.4\% & 12.2\% & 12.58\% \\
    5 &        4.6\%  &9.9\% & 11.4\% & 11.46\% \\
    6 &        4.3\%  &8.7\% & 10.2\% & 10.74\% \\
    7 &        3.8\%  &8.1\% & 9.2\% & 8.68\% \\
    8 &        2.8\%  &4.3\% & 7.4\% & 7.35\% \\
    9 &        2.6\%  &2.7\% & 1.8\% & 1.47\% \\
   10 &        2.5\%  &2.5\% & 1.2\% & 0.73\% \\
\bottomrule
\end{tabular}
\caption{The hashing power distribution for Ethereum, Bitcoin, Litecoin and Dogecoin as of~April~$2021$.}
\label{tab:hashing_power}
\end{table}

\begin{table}[]
\centering
\small
\begin{tabular}{l|llllllll}
\toprule
\bf Pct \% & 0\% & 10\% & 33\% & 50\% & 67\% & 90\% & 100\%                                       \\
\midrule
 \cite{kim2018measuring}                     & - & 99   & 151  & 208  & 231  & 285   & -  \\
 \cite{gencer2018decentralization}           & - & 92   & 125  & 152  & 200  & 276   & -  \\
 This work                                   & 0 & 95.5 & 138  & 180  & 215.5& 280/5 & 300 \\
\bottomrule
\end{tabular}
\caption{We base the latency distribution(ms) in this work on the mean statistics of the Ethereum P2P network provided by related works~\cite{kim2018measuring, gencer2018decentralization}.}
\label{tab:network_statistics}
\end{table}

\section{Sub-optimal two-points arbitrage for $N$-\aamm}

Algorithm~\ref{alg:suboptimal_arbitrage} shows the sub-optimal \ArbitrageForY strategy among $N+1$ \amms on $X/Y$ market.

\begin{algorithm}[]
\footnote
\SetAlgoLined
$M[0 \ldots N] \leftarrow$ AMMs with ascending $Y/X$ price\;
$l \leftarrow 0$ ; $r \leftarrow N$ \;
\While{True}{
    \tcp{Aggregation}
    $M_L \leftarrow $ aggregate exchanges $M[0]$ to $M[l]$\;
    $M_R \leftarrow $ aggregate exchanges $M[r]$ to $M[N]$\;
    \eIf{arbitrage between$M_L$ and $M_R$ is profitable?}{
        \tcp{Arbitrage computation}
        Simulate arbitrage between $M_L$ and $M_R$\;
        $p_{M_L} \leftarrow$ price of $M_L$ after arbitrage simulation\;
        $p_{M_R} \leftarrow$ price of $M_R$ after arbitrage simulation\;
        $b^{\text{shift}}_{\text{L}} \leftarrow (p_{M_L} > p_{M[l+1]}) \land ((l+1) < r)$\;
        $b^{\text{shift}}_{\text{R}} \leftarrow (p_{M_R} > p_{M[r-1]}) \land (l < (r-1))$\;
        \tcp{Synchronize volume}
        \If{$b^{\text{shift}}_{\text{L}}$}{
            $\delta_x^l \leftarrow$ such that $p_{M_L} == p_{M[l+1]}$ if $\ArbitrageForY(\delta_x^l, M[l], M[r])$ is executed\;
        }
        \If{$b^{\text{shift}}_{\text{R}}$}{
            $\delta_x^r \leftarrow$ such that $p_{M_R} == p_{M[r-1]}$ if $\ArbitrageForY(\delta_x^r, M[l], M[r])$ is executed\;
        }       
        \tcp{Swap execution}
        \If{$b^{\text{shift}}_{\text{L}}$ and $\delta_x^l \leq \delta_x^r$}{
            $\ArbitrageForY(\delta_x^l, M[l], M[r])$\;
            continue\;
        }
        \If{$b^{\text{shift}}_{\text{R}}$ and $\delta_x^l \geq \delta_x^r$}{
            $\ArbitrageForY(\delta_x^r, M[l], M[r])$\;
            continue\;
        }
    } {
        break\;
    }
}
\caption{Sub-optimal \ArbitrageForY strategy among $N+1$ \amms on $X/Y$ market. Our strategy iteratively performs three steps to extract arbitrage revenue, namely \textit{(i)} \amms aggregation; \textit{(ii)} arbitrage computation; \textit{(iii)} threshold computation; \textit{(iv)} swap execution, until the price of all \amms are leveled.}
\label{alg:suboptimal_arbitrage}
\end{algorithm}

\section{Past Arbitrage Volume Opportunities}
\label{sec:arbitrage_heuristics}
In the following, we quantify the volume and number of transactions performing two-point arbitrage on past blockchain data from block~$10,794,261$ (4th~September,~2020, Sushiswap's deployment) to block~$12,000,000$ (8th~March,~2021) ($186$ days).

\point{Arbitrage Heuristics} We adjust the heuristics proposed by Qin~\etal~\cite{qin2021quantifying} to detect past extracted two-point arbitrages. Recall that every \ArbitrageForX consists of two state transitions (cf.\ Equation~\ref{eq:AMM_arbitrage}). In the following we denote these two transitions as $\TransactXY(\delta^1_x, \delta^1_y)$ and $\TransactYX(\delta^2_y, \delta^2_x)$.

\begin{description}
\item[Heuristic 1] We assume that the arbitrageurs attempt to minimize their risks, and therefore execute both $\TransactXY$ and $\TransactYX$ atomically in the same transaction. Note that, unlike previous work~\cite{qin2021quantifying}, we do not constrain the execution order of \TransactXY and \TransactYX.
\item[Heuristic 2] The output ($\delta^1_y$) of \TransactXY must be greater than the input ($\delta^2_y$) of \TransactYX.
\item[Heuristic 3] The input ($\delta^1_x$) of \TransactXY must be less than the output ($\delta^2_x$) of \TransactYX. This implicitly assumes that an arbitrageur asserts a positive revenue in the smart contract, and reverts otherwise (ignoring the transaction fees).
\end{description}

\point{Empirical Results} We consider all assets and markets on Uni- and Sushiswap over~$186$ days, from block~$10,794,261$ (4th~September,~2020, Sushiswap's deployment) to block~$12,000,000$ (8th~March,~2021). Among the~$22,232,144$ Uni/Sushiswap related transactions, we identify a total of~$164,345$ ($0.7\%$) successful two-point arbitrage trades. These arbitrage activities realize a revenue of~$28,956$ ETH ($25,541,382$ DAI), contributing~$1.8B$ USD of trading volume to each of the two exchanges ($1.74\%$ and~$5\%$ of Uni- and Sushiswap's total transaction volume, respectively).

To better understand these arbitrage revenues, we deduct the transaction fees (gas costs) from these transactions. We find that $131,896$ ($80.07\%$) of the arbitrages are profitable, paying a total,~$4,348.11$ ETH ($32$ billion gas) in transaction fees. On average, two-point arbitrage yields an average revenue of~$118,626$ and~$25,730$ USD per day for arbitrageurs and miners, respectively.

\begin{table}[htb]
\footnotesize
\begin{tabular}{lllll}
\toprule
      & \multicolumn{2}{c}{\textbf{Trading Volume}} & \multicolumn{2}{c}{\textbf{Arbitrage Volume ($\%$)}} \\
\textbf{Month} & Uniswap &  Sushiswap & Uniswap  & Sushiswap \\
\midrule
20-09 &  12.2B &     2.2B & 215.4M(1.77\%) & 215.3M(9.73\%) \\
20-10 &   9.1B &   855.0M &  89.4M(0.99\%) & 89.2M(10.44\%) \\
20-11 &   9.6B &     2.0B & 117.8M(1.23\%) & 117.3M(5.92\%) \\
20-12 &  11.7B &     3.0B & 153.9M(1.31\%) & 151.6M(5.01\%) \\
20-01 &  25.3B &    11.7B & 452.7M(1.79\%) & 442.2M(3.77\%) \\
20-02 &  32.5B &    14.2B & 730.9M(2.25\%) & 724.1M(5.11\%) \\
\bottomrule
\end{tabular}
\caption{Trading volume, two-point arbitrage on Uniswap, Sushiswap. For example, in October~2020 arbitrage activities amount to a total volume of $855.0M$~USD, $10.44\%$ of the entire trading volume on Sushiswap ($14.2B$~USD).}
\end{table}

\point{Heuristic Limitations}
Heuristic 1 assumes that arbitrageurs perform arbitrage only within atomic transactions, as in to minimize execution risks. Naturally, some arbitrageurs may not perform atomic arbitrage, especially when not colluding with miners. Heuristic $1$ can therefore introduce false negatives by not capturing seemingly riskier arbitrage. Heuristics $2$ and $3$ will only detect arbitrage where asset $X$ increases but asset $Y$ does not decrease. While we lack quantitative insights on the heuristic accuracy, we choose to introduce relatively strict heuristics to reduce our false positive rate at the cost of underestimating the overall arbitrage transactions.
\end{appendices}

\end{document}